\documentclass[aps,pra,twocolumn,superscriptaddress]{revtex4-1}

\usepackage{amsmath}
\usepackage{graphicx}
\usepackage{hyperref}
\hypersetup{
    colorlinks=true,       
    linkcolor=blue,          
    citecolor=green,       
    filecolor=magenta,      
    urlcolor=blue          
}

\usepackage[dvipsnames]{xcolor}
\usepackage{tikz}

\definecolor{blue1}{rgb}{0,0.191,0.255}

\newcommand{\Rb}{$^{87}\text{Rb}$ }

\DeclareMathOperator{\Tr}{Tr}

\begin{document}

\preprint{}

\title{Dissipation induced dipole blockade and anti-blockade in driven Rydberg systems}


\author{Jeremy T. Young}
\affiliation{Joint Quantum Institute, NIST/University of Maryland, College Park, Maryland 20742 USA}

\author{Thomas Boulier}
\affiliation{Joint Quantum Institute, NIST/University of Maryland, College Park, Maryland 20742 USA}
\affiliation{Laboratoire Charles Fabry, Institut d'Optique Graduate School, CNRS, Universit\'{e} Paris-Saclay, 91127 Palaiseau cedex, France}

\author{Eric Magnan}
\affiliation{Joint Quantum Institute, NIST/University of Maryland, College Park, Maryland 20742 USA}
\affiliation{Laboratoire Charles Fabry, Institut d'Optique Graduate School, CNRS, Universit\'{e} Paris-Saclay, 91127 Palaiseau cedex, France}

\author{Elizabeth A. Goldschmidt}
\affiliation{United States Army Research Laboratory, Adelphi, Maryland 20783 USA}

\author{Ryan M. Wilson}
\affiliation{Department of Physics, The United States Naval Academy, Annapolis, MD 21402 USA}

\author{Steven L. Rolston}
\affiliation{Joint Quantum Institute, NIST/University of Maryland, College Park, Maryland 20742 USA}

\author{James V. Porto}
\affiliation{Joint Quantum Institute, NIST/University of Maryland, College Park, Maryland 20742 USA}

\author{Alexey V. Gorshkov}
\affiliation{Joint Quantum Institute, NIST/University of Maryland, College Park, Maryland 20742 USA}
\affiliation{Joint Center for Quantum Information and Computer Science, NIST/University of Maryland, College Park, Maryland 20742 USA}

\date{\today}

\begin{abstract}
We study theoretically and experimentally the competing blockade and anti-blockade effects induced by spontaneously generated contaminant Rydberg atoms in driven Rydberg systems. These contaminant atoms provide a source of strong dipole-dipole interactions and play a crucial role in the system's behavior. We study this problem theoretically using two different approaches. The first is a cumulant expansion approximation, in which we ignore third-order and higher connected correlations. Using this approach for the case of resonant drive, a many-body blockade radius picture arises, and we find qualitative agreement with previous experimental results. We further predict that as the atomic density is increased, the Rydberg population's dependence on Rabi frequency will transition from quadratic to linear dependence at lower Rabi frequencies. We study this behavior experimentally by observing this crossover at two different atomic densities. We confirm that the larger density system has a smaller crossover Rabi frequency than the smaller density system. The second theoretical approach is a set of phenomenological inhomogeneous rate equations. We compare the results of our rate equation model to the experimental observations in [E. A. Goldschmidt, et al., PRL \textbf{116}, 113001 (2016)] and find that these rate equations provide quantitatively good scaling behavior of the steady-state Rydberg population for both resonant and off-resonant drive.
\end{abstract}

\pacs{}

\maketitle

\section{Introduction}

Ultracold atomic systems are an ideal setting for studying many-body quantum systems due to their large degree of control and tunability. Rydberg atoms in particular are a key ingredient in many of these systems, primarily for their strong, long-range interactions and long lifetimes  \cite{Saffman2010b,Pohl2011}. Because of these features, the possibilities Rydberg atoms provide are incredibly diverse, including simulating many-body driven-dissipative systems \cite{Malossi2014,Lee2011,Honing2013,Lee2013,Lee2012a}, simulating topological states of matter \cite{Dauphin2012,vanBijnen2015}, and applications in quantum information \cite{Saffman2010b,Xia2015,Weimer2009a}. One aspect of several of these systems is Rydberg dressing \cite{Henkel2010,Jau2016,Dauphin2012,vanBijnen2015,Glaetzle2015,Pupillo2010a, Glaetzle2014b}, which provides a means of creating soft-core potentials and is achieved by weakly dressing a ground state with a Rydberg state \cite{Honer2010,Johnson2010,Helmrich2015,Glaetzle2012a}. However, recently it has been found that through spontaneous decay and blackbody radiation, nearby contaminant Rydberg states can become populated and can drastically modify the system's behavior via the resultant dipole-dipole interactions \cite{Goldschmidt2016,Aman2016,Boulier2017}. While Rydberg dressing has been achieved with up to 200 atoms \cite{Zeiher2016}, the possible appearance of contaminant states necessitates a form of post-selection, with far more post-selection required to increase the strength and range of the dressed potentials or to increase the system size. On the other hand, the manner in which the dipole-dipole interactions arise is unique. Rather than coherent processes (e.g. drive) leading to interactions, we instead have a system in which a dissipative process leads to interactions. As a result, this provides an interesting platform for studying driven-dissipative systems in which coherent processes both compete with and rely on dissipation, whereas they typically only compete in most Rydberg systems.

There are two primary mechanisms which lead to the broadening induced by the dipole-dipole interactions: blockade and anti-blockade. Blockading is the process in which a nominally resonant excitation becomes off-resonant due to interactions \cite{Urban2009}, which can lead to the formation of superatoms with collectively enhanced Rabi frequencies \cite{Tong2004,Heidemann2007,Honer2010}. Complementary to this, anti-blockading (also known as facilitated resonance) is the process in which a nominally off-resonant excitation becomes resonant due to interactions and plays an important role in phenomena such as Rydberg aggregation \cite{Urvoy2015,Schempp2014,Garttner2013}. Both of these mechanisms play a crucial role in all Rydberg systems, but most investigations have focused on $1/r^6$ diagonal van der Waals interactions. In such systems, when the drive is resonant, blockading dominates, while when the drive is off-resonant, anti-blockade often dominates. However, we are interested in $1/r^3$ off-diagonal (``flip-flop'') dipole-dipole interactions. As a result of the off-diagonal nature and angular dependence of dipole-dipole interactions, blockading and anti-blockading will behave qualitatively differently than for van der Waals interactions, with both effects competing with one another in complicated ways. This complicates any attempt to truncate the Hilbert space via blockading or dephasing, which has been successful in studying Rydberg systems with diagonal interactions \cite{Weimer2008, Low2009,Lesanovsky2013}.

\begin{figure}
\includegraphics[scale=.33]{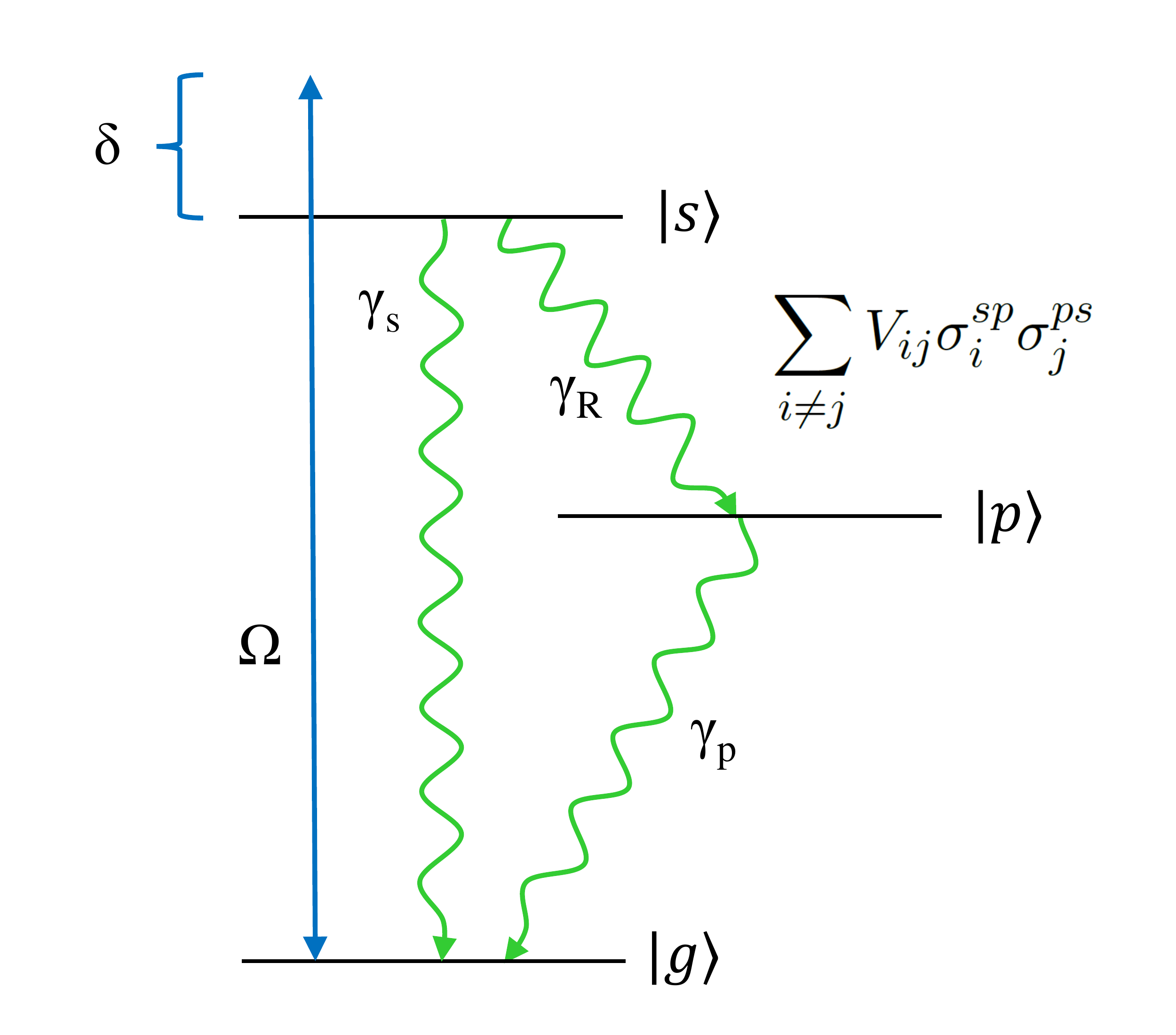}
\caption{Theoretical three-level system. The $g$ and $s$ states are coupled via a classical laser with Rabi frequency $\Omega$ and detuning $\delta$, while the $s$ and $p$ states interact via a dipole-dipole interaction $\sum_{i \neq j} \sigma_i^{sp} \sigma_j^{ps}$. There are three decay processes: $s \to g$, $p \to g$, and $s \to p$, with decay rates of $\gamma_s,\gamma_p,\gamma_R$ respectively. \label{levels}}
\end{figure}

In this paper, we study the steady states of a driven-dissipative model in which Rydberg dipole-dipole interactions are induced via dissipation as in Refs. \cite{Goldschmidt2016,Aman2016,Boulier2017}. In all three references, a ground state is driven to a Rydberg $s$ state with Rabi frequency $\Omega$ and detuning $\delta$. Through spontaneous decay and blackbody stimulated transitions, nearby (in principal quantum number) contaminant $p$ states are populated. These $p$ states interact strongly with subsequently driven $s$ states via dipole-dipole interactions, leading to strong dephasing. A simplified model of this is illustrated in Fig.~\ref{levels}.

We approach this system theoretically in two different ways. The first is by considering evolution under the full master equation and applying a cumulant expansion approximation, which allows for two-atom correlations but ignores higher-order correlations. This is motivated by the presence of dissipation, which causes high-order correlations to decay faster than low-order correlations, and allows the many-body problem to be treated numerically. This approach has previously been used in a variety of systems, including nonlinear optics \cite{Schack1990}, cavity quantum electrodynamics \cite{Meiser2010,Henschel2010}, and other driven-dissipative systems with similar interactions \cite{Zhu2015, Chan2015a}. The second is a set of phenomenological inhomogeneous rate equations in which the decoherence strength for a given atom is determined by the population and interaction strength of neighboring contaminant states. Similar types of rate equations have been considered previously in other Rydberg systems \cite{Aman2016,Lesanovsky2013,Letscher2017a, Letscher2017b}.

For the cumulant expansion approach, we restrict our focus to the case of resonant drive ($\delta = 0$), and we consider both one-dimensional (1D) and three-dimensional (3D) systems. We find that in spite of the angular dependence and flip-flop nature of dipole-dipole interactions, a blockade radius interpretation still arises. However, the many-body blockade radius is found to be smaller than and to behave qualitatively differently from the two-body blockade radius. This occurs due to an interplay between both blockade and anti-blockade effects. Additionally, the steady-state Rydberg population exhibits power law decay over several orders of magnitude as a function of interaction strength, although the decrease in population is not as pronounced as observed experimentally in Ref. \cite{Goldschmidt2016}, which is possibly due to the importance of higher-order correlations and many-body effects. Finally, we observe at high Rabi frequencies a trend away from the expected quadratic dependence of the Rydberg population on Rabi frequency. For higher atomic densities, this trend occurs at lower Rabi frequencies. One reason to expect this is that at sufficiently low Rabi frequency, the density of Rydberg atoms becomes small and dipole-dipole interactions become irrelevant. We verify this experimentally by studying the low Rabi frequency behavior at two different densities. While there are still a number of qualitative and quantitative differences with theory, we find that the crossover occurs at lower Rabi frequencies for higher densities as expected. Furthermore, even when the scaling behavior is quadratic, the experimentally observed Rydberg populations are still much smaller than expected from single-particle physics, indicating that interactions still play an important role in this regime.

For the rate equation approach, we consider both resonant ($\delta = 0$) and off-resonant ($\delta \neq 0$) drive in 3D. While van der Waals interactions are diagonal and can be thought of as leading to an effective detuning, dipole-dipole interactions are off-diagonal and cannot be thought of in the same way. Therefore, we treat them as a source of decoherence, as the contribution to an effective detuning will depend strongly on the spatial configuration of the atoms and cannot be simply represented by a single value. Since stronger interactions will have a larger effect, we make this decoherence strength proportional to the interaction strength and population of the contaminant states. Finally, we focus on inhomogeneous rate equations to reflect the inherent inhomogeneity in the system due to spontaneous decay. We find that such an approach accurately captures the experimentally observed behavior of the Rydberg population in Ref. \cite{Goldschmidt2016}, both on resonance and off resonance. Furthermore, the exact details of how the decoherence is implemented primarily affects the Rydberg population lineshapes, while the qualitative scaling behavior remains unchanged. However, this model fails to accurately capture both the early time and low Rabi frequency behavior. In these regimes the number of contaminant atoms is small and individual Rydberg atoms can affect the system more easily, so the spatial configuration of the atoms, and thus their correlations, play a more important role.

The remainder of the paper is organized in the following manner. In Sec.~\ref{modelsec}, we describe our theoretical approaches to this system, including the details of the cumulant expansion approximation and our phenomenological inhomogeneous rate equations. In Sec.~\ref{cumulantressec}, we present the theoretical results of the cumulant expansion approximation as well as an experimental examination of the crossover from quadratic to linear dependence of the Rydberg population on Rabi frequency. In Sec.~\ref{ratesec}, we present the theoretical results of our phenomenological inhomogeneous rate equations and compare them to the experimental results of Ref. \cite{Goldschmidt2016}. Finally, in the Appendices, we include several details omitted from the main text.

\section{Theoretical Models}
\label{modelsec}

In order to study the effect of contaminant $p$ states on driven-dissipative Rydberg systems, we consider a three-level system composed of states $|g\rangle$, $|s\rangle$, and $|p\rangle$, corresponding to the ground, $ns$, and $mp$ states, where $n$ and $m$ are the principal quantum numbers of the $s$ and $p$ Rydberg states, respectively. Although there are generally multiple $mp$ states with large enough dipole matrix elements with the $ns$ state to affect the dynamics of the system, we consider here only one contaminant $p$ state for simplicity. We also assume a nonzero magnetic field, as is the case in Refs. \cite{Goldschmidt2016,Aman2016,Boulier2017}. Our effective three-level model is illustrated in Fig.~\ref{levels}.

The transition between $|g \rangle$ and $|s \rangle$ is driven via a classical laser with Rabi frequency $\Omega$ and detuning $\delta$, where we have chosen to define our Rabi frequency as half of the traditional definition in order to avoid carrying around extra factors of two. Additionally, the $|s\rangle $ and $|p \rangle$ states will interact according to a flip-flop dipole-dipole interaction. While van der Waals interactions are typically present, we ignore them here since they are weak compared to the dipole-dipole interactions that we want to study. Together, these result in the following Hamiltonian

\begin{equation}
H =  \sum_i \left[ - \delta \sigma_i^{ss} + \Omega (\sigma_i^{gs} + \sigma_i^{sg})\right] + \sum_{i \neq j} V_{ij} \sigma_i^{sp} \sigma_j^{ps} ,
\end{equation}
where we define operators $\sigma^{\alpha \beta}_i= |\alpha \rangle_i \langle \beta|_i$. The last sum is over both $i$ and $j$. The interaction strength between atoms $i$ and $j$ is given by
\begin{equation}
V_{ij} = \frac{C_3}{r_{ij}^3}(1-3\cos^2 \theta_{ij}),
\end{equation}
where $C_3$ defines the strength of the dipole-dipole interactions, $r_{ij}$ is the separation between atoms $i$ and $j$, and $\theta_{ij}$ is the angle the displacement vector $\mathbf{r}_{ij}$ makes with the quantization axis, which is determined by the magnetic field. While there are dipole-dipole interactions which have a different angular dependence, they are off-resonant due to Zeeman shifts from the magnetic field and are less relevant as a result. However, for sufficiently strong interactions relative to the Zeeman shifts, these interactions could potentially play a more important role. 
 
In addition to Hamiltonian dynamics, decay also plays a crucial role as the source of the contaminant $p$ states. We treat both aspects according to a master equation description
\begin{equation}
\dot{\rho} = -i [H, \rho] + \gamma_s \mathcal{L}_s[\rho]+ \gamma_p \mathcal{L}_p[\rho]+ \gamma_R \mathcal{L}_R[\rho],
\end{equation}
where $\gamma_s$, $\gamma_p$, and $\gamma_R$ are the decay rates from $|s \rangle$ to $|g \rangle$, $|p \rangle$ to $|g \rangle$, and $|s \rangle$ to $|p \rangle$ respectively. We ignore blackbody radiation from the $p$ state back to the $s$ state since most of the corresponding blackbody radiation goes to other $s$ and $d$ states. Throughout this paper, we will take $\gamma_s = \gamma_p =1$ and $\gamma_R = .3$, which provides comparable branching ratios to Ref. \cite{Goldschmidt2016}, although this comparison is complicated by the fact that there are many relevant $p$ states as well as decay to non-participating ground states. The associated Lindblad operators for decay $\mathcal{L}_s$, $\mathcal{L}_p$, and $\mathcal{L}_R$ are given below

\begin{subequations}
\begin{equation}
\mathcal{L}_s[\rho]= \sum_i \left[\sigma_i^{gs} \rho \sigma_i^{sg} - \frac{1}{2} \{\rho,\sigma_i^{ss}\}\right],
\end{equation}
\begin{equation}
\mathcal{L}_p[\rho]= \sum_i \left[\sigma_i^{gp} \rho \sigma_i^{pg} - \frac{1}{2} \{\rho,\sigma_i^{pp}\}\right],
\end{equation}
\begin{equation}
\mathcal{L}_R[\rho]= \sum_i \left[\sigma_i^{ps} \rho \sigma_i^{sp} - \frac{1}{2} \{\rho,\sigma_i^{ss}\}\right].
\end{equation}
\end{subequations}

We are most interested in the steady state of the above master equation. However, this can only be determined numerically for up to approximately ten atoms, far from any sort of long-range many-body behavior we are interested in. One common approach to this problem is to use Gutzwiller mean field theory, which ignores the effects of correlations and assumes the steady-state density matrix is a product state \cite{Rokhsar1991, Diehl2010}. In Appendix \ref{MFT}, we explain why this technique fails to capture the behavior of our model. Instead, we will approach the problem via a cumulant expansion approximation, which we discuss below.

\subsection{Cumulant Expansion}
\label{cumulantsec}

Rather than truncate the hierarchy of differential equations at the level of single atom operators as in Gutzwiller mean field theory, we instead use a second-order cumulant expansion approximation, which continues one step further and allows for correlations between pairs of atoms \cite{Schack1990,Meiser2010,Henschel2010,Zhu2015}. Formally, this amounts to making the following approximation
\begin{equation}
\begin{aligned}
\langle \mathcal{A}_i \mathcal{B}_j \mathcal{C}_k \rangle =& \ \langle \mathcal{A}_i \mathcal{B}_j \rangle \langle \mathcal{C}_k \rangle + \langle \mathcal{C}_k \mathcal{A}_i \rangle \langle \mathcal{B}_j \rangle \\& + \langle \mathcal{B}_j \mathcal{C}_k \rangle \langle \mathcal{A}_i \rangle - 2 \langle \mathcal{A}_i \rangle \langle \mathcal{B}_j \rangle \langle \mathcal{C}_k \rangle, 
\end{aligned}
\end{equation}
where $i,j,k$ correspond to distinct atoms and $\mathcal{A}, \mathcal{B}, \mathcal{C}$ are single atom operators ($\sigma^{\alpha \beta}$ in our model). This is equivalent to setting all three-atom and higher connected correlations to zero. The $n$th-order connected correlation accounts for inherently $n$-body correlations which cannot be understood in terms of lower-order correlations. This approximation is justified under the assumption that two-atom correlations will dominate, which is often the case when dissipation and decoherence are involved. This truncation reduces a set of $\sim 9^N$ equations to a set of $\mathcal{O}(N^2)$ equations, where $N$ is the number of atoms. 

Restricting our focus to a lattice with unit filling, we may use translational symmetry and truncate correlations past a certain distance (where they are negligible) in order to reduce this further to a set of $\mathcal{O}(M)$ equations, where $M$ is the number of displacement vectors considered. For 3D, we take all correlations involving distances greater than 16 times the lattice spacing to be zero. For 1D, we choose this distance to be 100 times the lattice spacing. Furthermore, we take advantage of the four different reflection symmetries present in the dipole interaction in 3D, reducing the number of nonlinear coupled ordinary differential equations by a further factor of 16. Finally, there is also a $U(1)$ symmetry present in the form of $|p \rangle \to e^{i \phi} |p \rangle$, which forces some terms in the density matrix to be zero in steady state. Since we are assuming correlations past a certain distance to be negligible, we restrict the strength of $C_3$ so that the interaction strengths beyond this distance are not large compared to the decay rates. By using these symmetries, we are able to consider large system sizes and, correspondingly, large interaction strengths. Steady-state behavior is found by numerically integrating the resultant effective equations of motion using a 4th-order Runge-Kutta method. Examples of the resultant effective nonlinear equations of motion are given in Appendix \ref{cumulant}.

To understand whether we can expect the cumulant expansion to give reasonably accurate results, we use a quantum trajectories approach \cite{Dalibard1992,Dum1992,Plenio1998,Daley2014} to compare the approximate cumulant expansion with exact numerics of quantum trajectories for small system sizes and find that they produce results that are similar in this limit. While the rest of this paper focuses on parameter regimes well outside this limit, this demonstrates that this approximation can capture the effects of the interactions. The results of this comparison are covered in detail in Appendix \ref{traj}.

\subsection{Inhomogeneous Rate Equations}

In addition to the cumulant expansion approach on resonance, we also study a set of phenomenological inhomogeneous rate equations. The fundamental assumption we make in forming these rate equations is that rather than an effective shift in the detuning of individual sites, nearby $p$ atoms cause dephasing proportional to their interaction strength. This is motivated by the fact that the dipole-dipole interactions are off-diagonal, so their effect cannot be strictly understood in terms of effective detunings. Additionally, we take these rate equations to be spatially inhomogeneous by considering atoms which are independently and identically distributed according to a 3D Gaussian probability distribution. This is done to capture the fact that in a real system, the spontaneous decay will lead to a spatially inhomogeneous distribution of $p$ atoms. These assumptions lead to the following set of rate equations
\begin{subequations}
\label{rate1}
\begin{equation}
\dot{s_i} = R_i(g_i - s_i) - (\gamma_s + \gamma_R) s_i,
\end{equation}
\begin{equation}
\dot{p_i} =  \gamma_R s_i - \gamma_p p_i,
\end{equation}
\begin{equation}
\dot{g_i} = -R_i(g_i - s_i) + \gamma_s s_i + \gamma_p p_i,
\end{equation}
\end{subequations}
where the pumping rate $R_i$ is given by
\begin{equation}
\label{rate2}
R_i = \frac{\Omega^2}{\delta^2 + \Gamma_i^2/4}\Gamma_i,
\end{equation}
and the dephasing rate $\Gamma_i$ is given by
\begin{equation}
\label{rate3}
\Gamma_i = \gamma_s + \gamma_R + C_3 \left|\sum_{j \neq i} V_{ij} p_j\right|.
\end{equation}
The variables $s_i, p_i, g_i$ refer respectively to the $s, p, g$ populations at site $i$.

One important feature of these rate equations is that the scaling behavior of the steady state population is generally insensitive to the exact manner in which the interactions are included in the dephasing rate, and several different choices produce the observed experimental scaling. They primarily differ in the coefficient of the linewidth and of the resonant Rydberg population scaling as well as the resultant lineshapes. For example, Ref. \cite{Boulier2017} considers a set of homogeneous rate equations with $\Gamma = \gamma_s + \gamma_R + n_{\textrm{3D}} C_3 p$, where $n_{\textrm{3D}}$ ($n_{\textrm{1D}}$) is the density of atoms in 3D (1D). This model captures many features of the width behavior, but it predicts dome-like lineshapes rather than the experimentally observed Lorentzian lineshapes. Similarly, the spatial distribution of atoms in these types of models can also affect the lineshape, with a lattice distribution often leading to more dome-like lineshapes in general. Our choice of effective dephasing is the simplest choice we have found which results in near-Lorentzian lineshapes.

\section{Cumulant Expansion Results}
\label{cumulantressec}

\subsection{Divergences}

\begin{figure*}
\includegraphics[scale=.32]{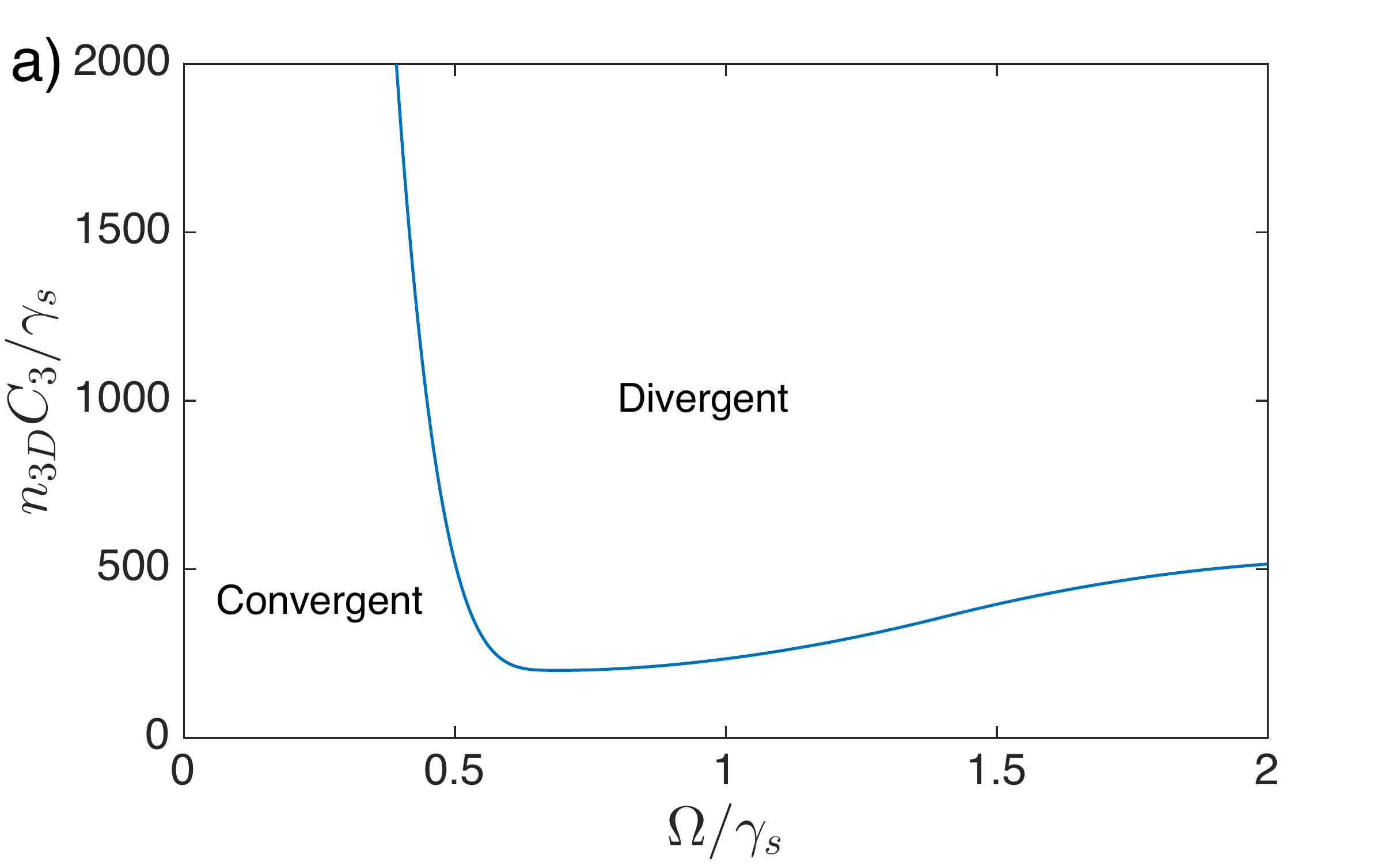}
\includegraphics[scale=.32]{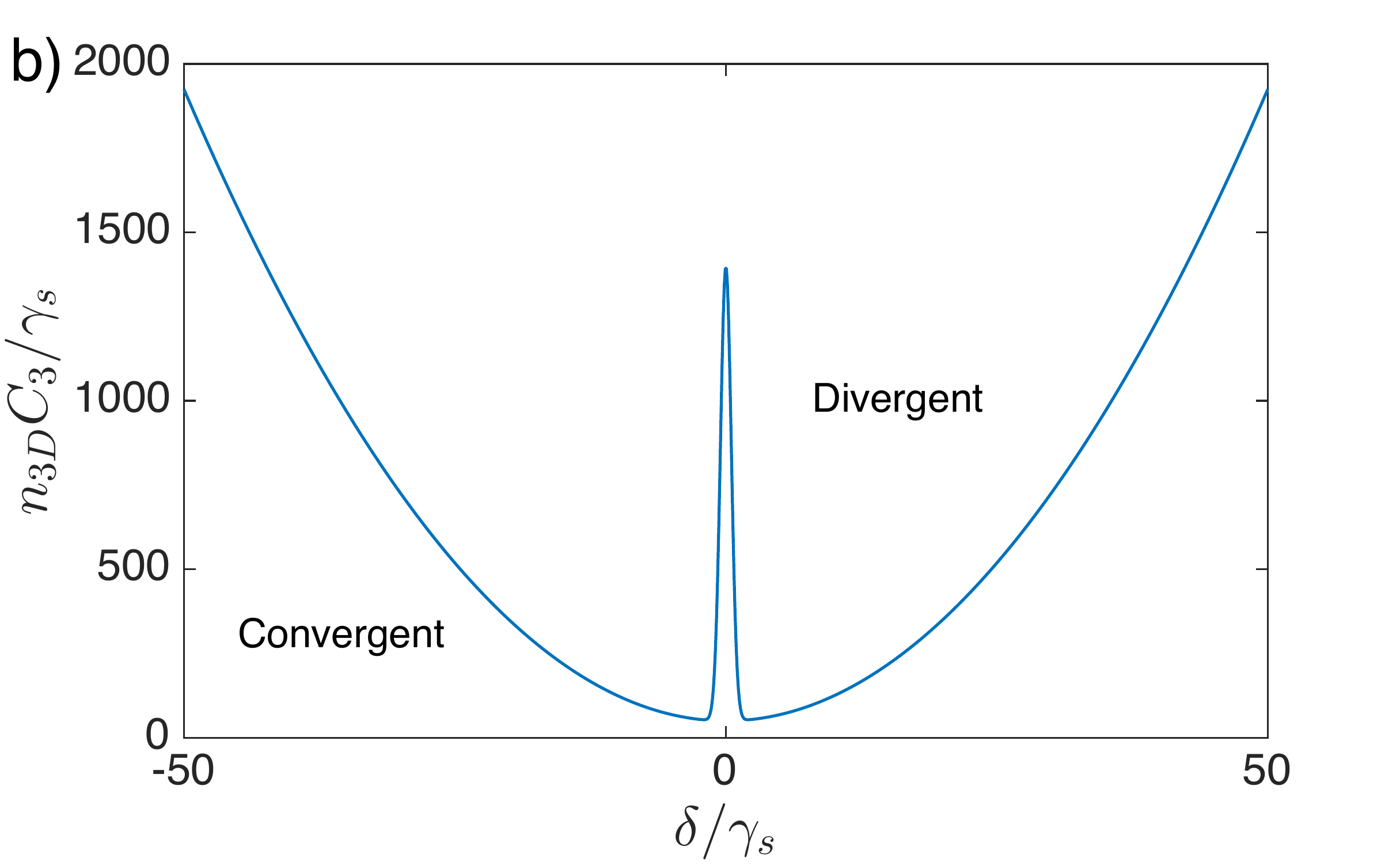}
\caption{Approximate divergence diagrams for cumulant expansion in 3D. (a) Divergence diagram on resonance ($\delta = 0$). Most of the low Rabi frequencies on resonance are convergent, although they become less stable as the interactions are increased. Sufficiently large Rabi frequencies are also convergent, where steady-state populations begin to saturate. (b) Divergence diagram for $\Omega/\gamma_s = .4$. A very narrow region near resonance is convergent for sufficiently small interaction strengths. The outer edges of the divergent region grow approximately quadratically in detuning in this parameter regime.}
\label{divdiag}
\end{figure*}

In this section, we discuss the results of the cumulant expansion approximation, which takes third-order and higher connected correlations to be zero. One issue that can arise under this approximation is the presence of unphysical divergences. Although we focus our attention on resonance, these divergences generally occur at intermediate detunings and Rabi frequencies. Rather than divergences due to numerical error, these divergences appear to be fundamental instabilities of the nonlinear differential equations, where there is only a single, unstable steady state. Furthermore, these divergences do not appear to be present in 1D systems, but they are very relevant in 3D systems. The origin of these instabilities is most likely the importance of the higher-order correlations that we have ignored \cite{Schack1990}, although finite size effects could play a role as well. 

In Fig.~\ref{divdiag} we plot a diagram showing the approximate parameter regimes where the cumulant expansion leads to a divergence. The regions where one would expect high-order correlations to be more important are exactly those where the divergences are present. The vast majority of the data in Ref. \cite{Goldschmidt2016} is well into these divergent regions, with $C_3 n_{\textrm{3D}}/\gamma_s$ often an order of magnitude larger than what we can treat numerically. Interestingly, the outer edge of the divergent region appears to grow approximately as $\delta \sim \sqrt{C_3 n_{\textrm{3D}}/\gamma_s}$. This is exactly the experimentally observed scaling behavior of the linewidth as a function of interaction strength, so the observed scaling of the linewidths may be reflected in the behavior of the divergences. As the interaction strength is increased, the linewidth increases, expanding the region where high-order correlations are important. Thus if high-order correlations are the origin of the divergences, we would expect these divergences to grow in the same manner as the lineshapes themselves, which is exactly what we find.

In order to determine the Rydberg populations in divergent parameter regimes, we further consider two more terms in the master equation that represent decoherence on the $|s\rangle$ and $|p \rangle$ states.

\begin{subequations}
\begin{equation}
\mathcal{L}_d^s[\rho] = \gamma^s_d \sum_i \left[\sigma_i^{ss} \rho \sigma_i^{ss} - \frac{1}{2} \{\rho,\sigma_i^{ss}\}\right],
\end{equation}
\begin{equation}
\mathcal{L}_d^p[\rho] = \gamma^p_d \sum_i \left[\sigma_i^{pp} \rho \sigma_i^{pp} - \frac{1}{2} \{\rho,\sigma_i^{pp}\}\right],
\end{equation}
\end{subequations}
where $\gamma_d^s$ and $\gamma_d^p$ correspond to the strength of decoherence on $|s\rangle$ and $|p \rangle$, respectively. We set $\gamma_d^{s} = \gamma_d^{p} = \gamma_d$ for simplicity. In terms of the differential equations themselves, this amounts to including extra decay on the coherences but not on the populations. When a sufficient amount of decoherence is included, parameter regimes which were formerly divergent become convergent. This is consistent with the understanding that the instabilities are a result of the importance of higher-order correlations, since decoherence decreases correlations. We focus on the cases of resonant drive because they only require a small amount of extra decoherence to become convergent. The amount of decoherence necessary for convergence ($\gamma_d/\gamma_s \approx .1$) is small compared to the decay rates and certainly smaller than any potential experimental source of decoherence which we have not included in our model.

\begin{figure*}
\includegraphics[scale=.3]{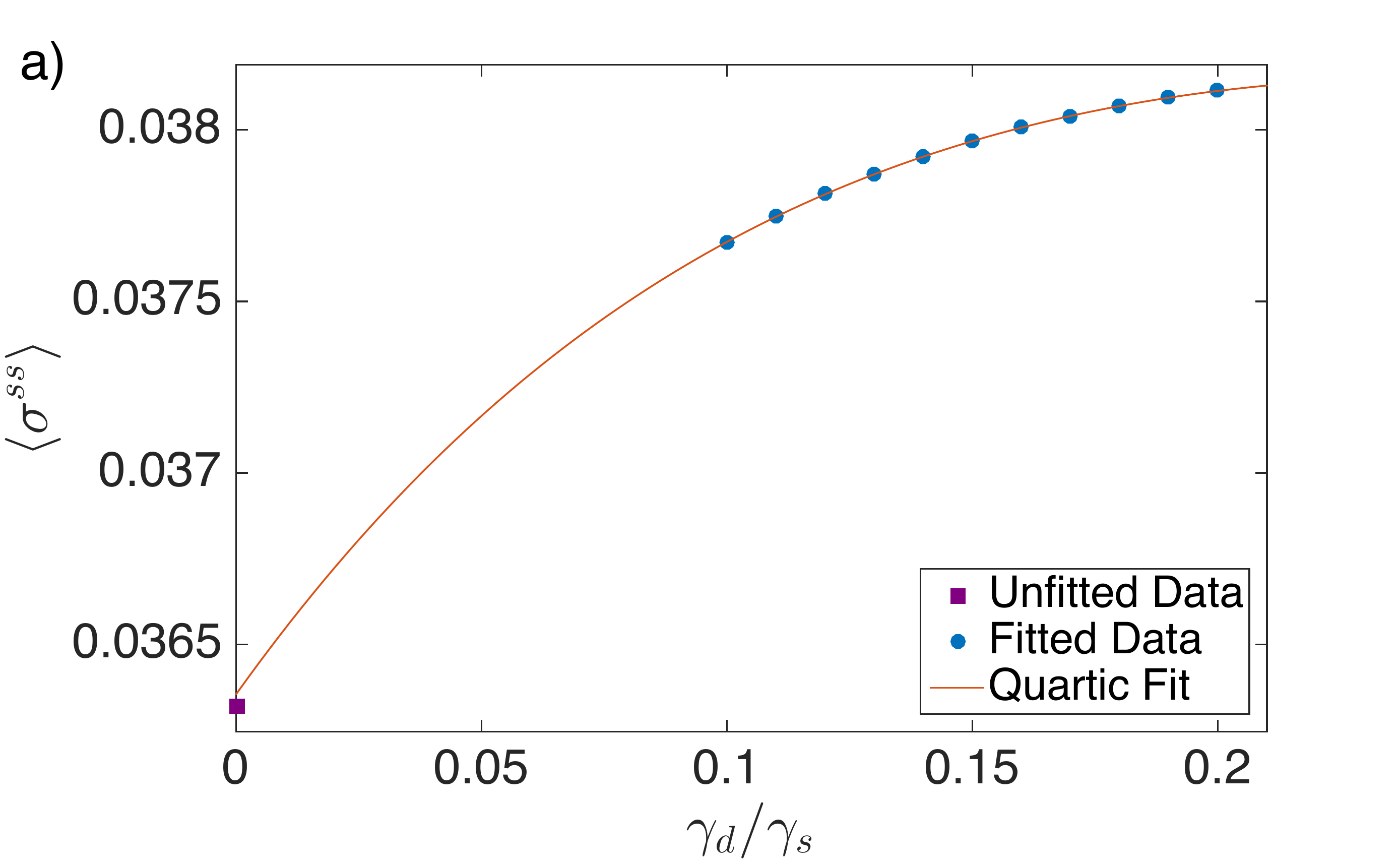}
\includegraphics[scale=.41]{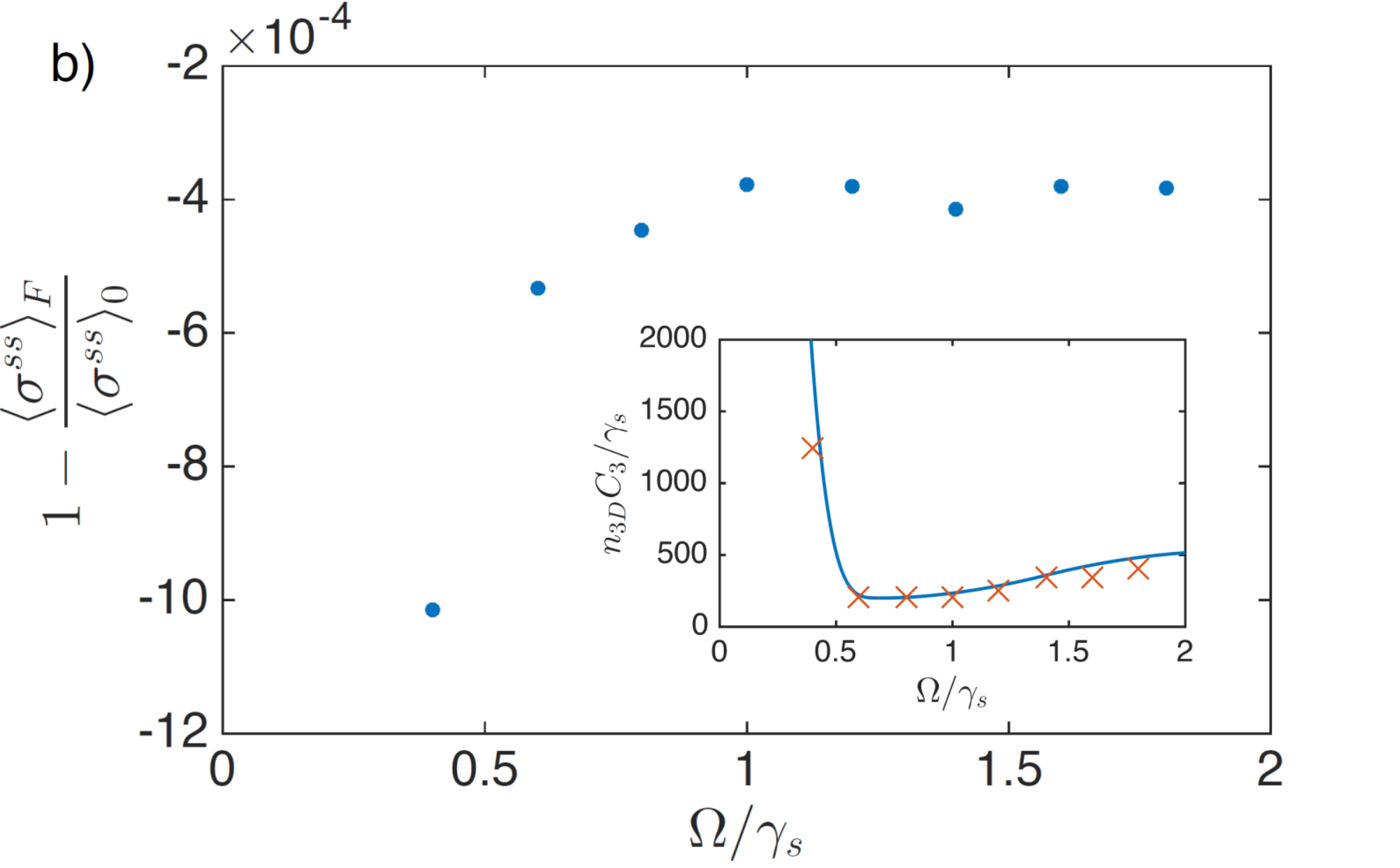}
\caption{Illustration of how fitting the Rydberg population as a function of decoherence is used to approximate the Rydberg population in divergent regions. (a) $n_{\textrm{3D}} C_3/\gamma_s = 1250 , \Omega/\gamma_s = .4$. The orange line corresponds to a quartic fit of the blue circles where $\gamma_d/\gamma_s > .1$. The purple square denotes the population for $\gamma_d = 0$, which is convergent for these parameters. (b) Relative error of Rydberg population extracted from a $\gamma_d/\gamma_s > .1$ fit compared to actual Rydbeg population at $\gamma_d = 0$, denoted $\langle \sigma^{ss} \rangle_{\textrm{F}}$ and $\langle \sigma^{ss} \rangle_0$ respectively, for several choices of parameters just outside of the divergent region. The inset shows the parameters used in the divergence diagram on resonance, where the orange $\times$'s denote the parameters used and the blue line separates the convergent and divergent regions, with the top right corresponding to the divergent region.}
\label{dec}
\end{figure*}

More importantly, the effect of increasing decoherence modifies the steady-state population in a simple way. As one crosses from the convergent region to the divergent region, the convergent steady state continuously becomes a divergent steady state. This provides a way to estimate the expected population when no decoherence is included. We achieve this by fitting the numerics for different decoherences and extrapolating the population at $\gamma_d = 0$ according to the fit. The accuracy of this technique is illustrated in Fig.~\ref{dec}, where we apply it to several choices of parameters parameters just outside of the divergent region. The populations we extrapolate from the fits differ from the actual populations at $\gamma_d$ by at most one tenth of a percent. However, as one moves far into the divergent regime, this method becomes increasingly less accurate because stronger decoherence is necessary for convergence.

\subsection{Blockade Radius Reduction}

\begin{figure*}
\includegraphics[scale=.3]{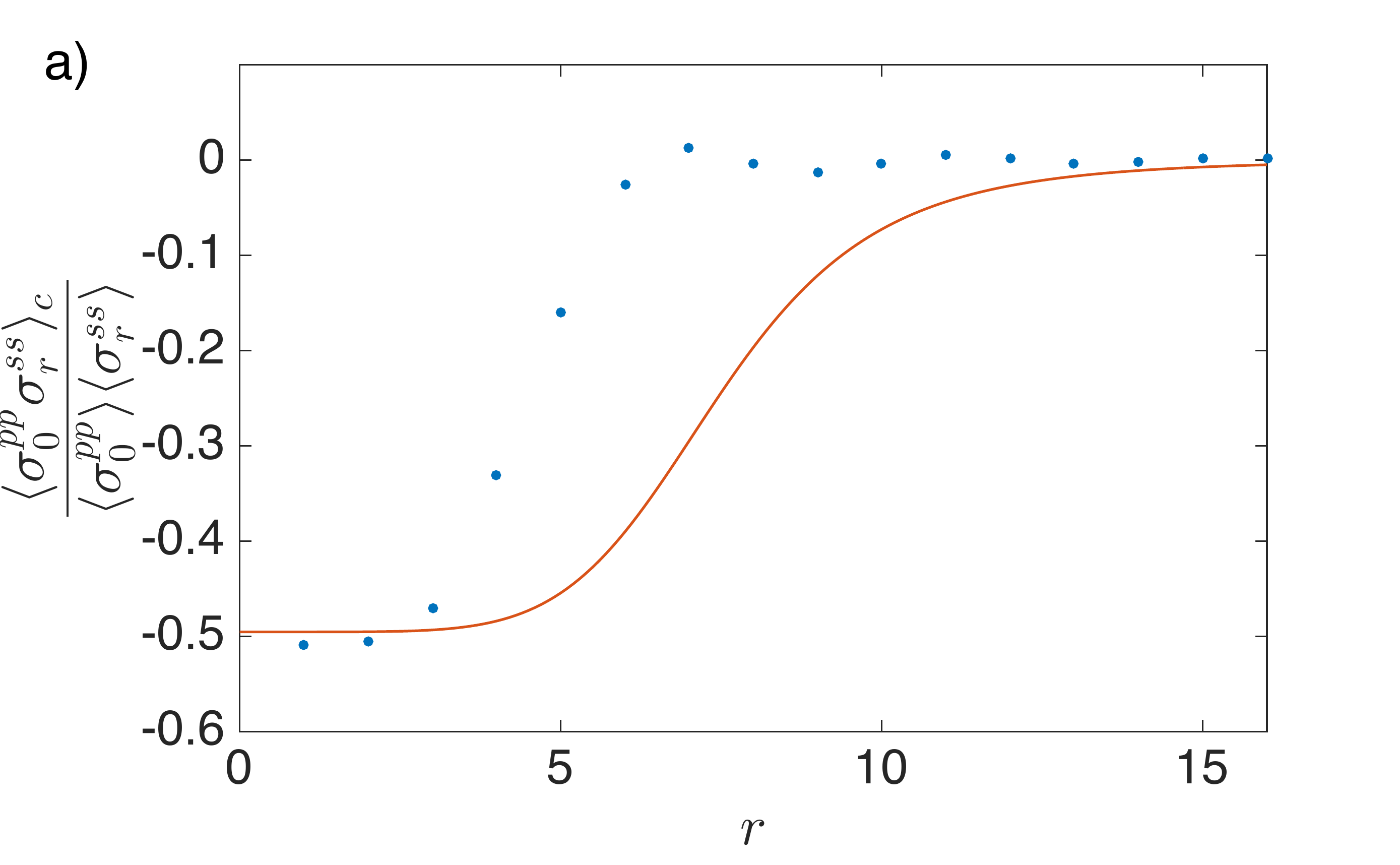}
\includegraphics[scale=.3]{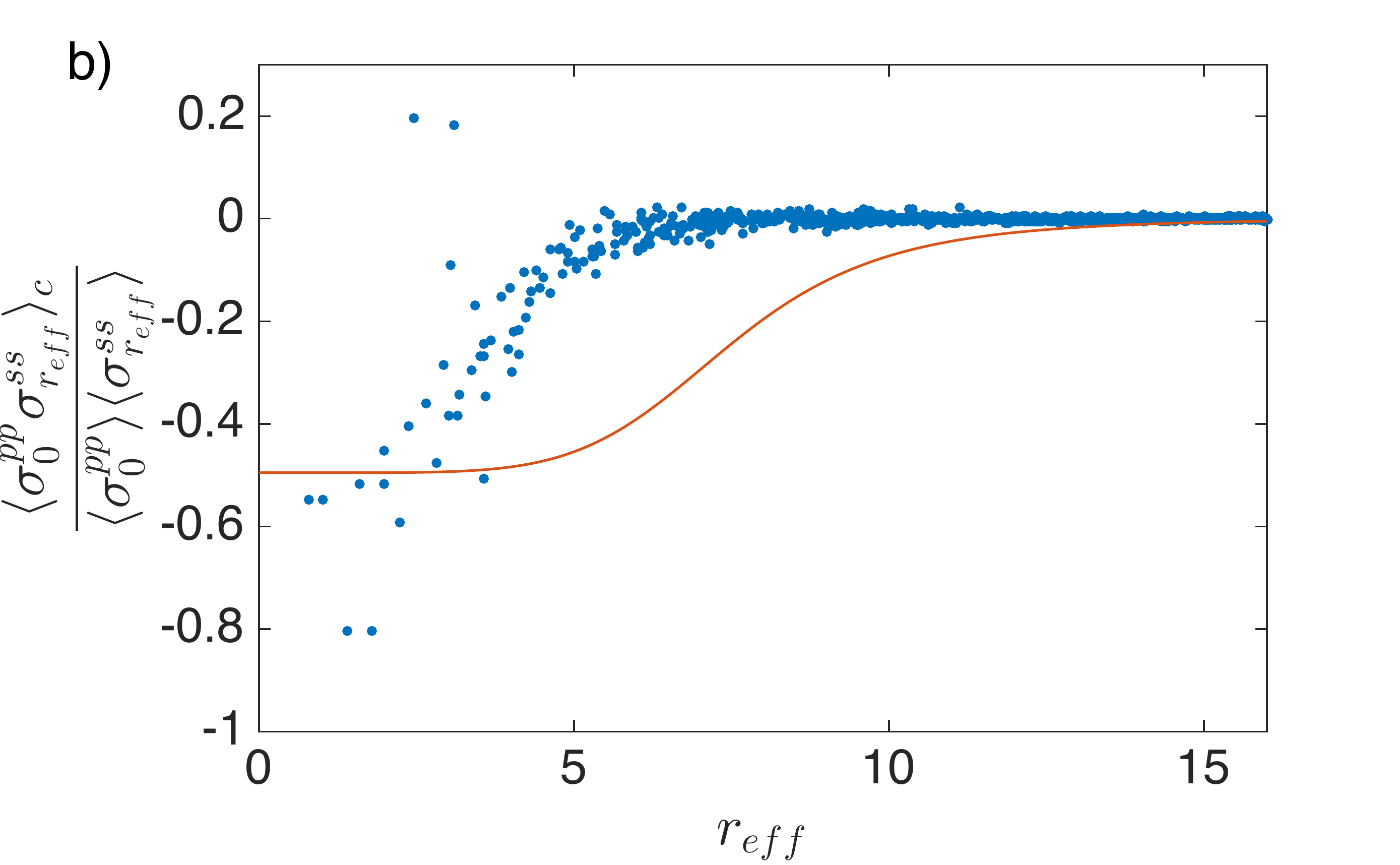}
\caption{Correlations between $s$ and $p$ atoms for $\Omega/\gamma_s = .4$. The blue dots are from the cumulant expansion while the orange line corresponds to exact calculations for just two atoms separated a distance $r$. These are plotted in (a) 1D for $n_{\textrm{1D}}^3 C_3/\gamma_s = 800$ and (b) 3D for $n_{\textrm{3D}} C_3/\gamma_s = 800$.}
\label{corr}
\end{figure*}

A concept that is often useful to consider in Rydberg systems is the blockade radius \cite{Urban2009, Heidemann2007, Tong2004}. Although in this case we are not considering the usual $1/r^6$ diagonal van der Waals interactions, the general effect of interactions suppressing excitations will occur in a similar fashion. However, due to the off-diagonal nature of the interactions, the effect of blockading will be modified in a non-trivial way in many-body systems. 

The blockade radius $r_b$ is often defined as the distance at which the interaction strength is equal to the effective Rabi frequency, $V(r_b) = \Omega_{\textrm{eff}}$. The effective Rabi frequency is defined self-consistently as $\Omega_\textrm{eff} = \sqrt{N_b} \Omega$, where $N_b$ is the number of atoms in a blockade volume. In this limit, where only one excitation is possible within a blockade volume, a superatom picture arises in which many atoms behave as an effective two-level atom \cite{Heidemann2007,Zeiher2015}. Since each Rydberg superatom blockades a volume of $V_b \propto r_b^3$, the total number of Rydberg atoms is proportional to $1/V_b$. One might na\"{i}vely expect to apply a similar analysis in the case of the contaminant p states, with each producing a large blockade volume in which $s$ atoms can no longer be excited or de-excited. However, were this the case, the Rydberg populations in Ref. \cite{Goldschmidt2016} would be much lower than observed because the long-range behavior of the dipole-dipole interaction corresponds to blockade volumes that are on the order of the system size, while the corresponding populations of Rydberg states can be in the hundreds. As a result, the size of the blockade volume due to the contaminant states, if one exists, must be significantly smaller in order to account for the observed Rydberg populations. 

In a many-body Rydberg system, individual atoms are affected by interactions due to multiple atoms. In the case of diagonal van der Waals blockading, these interactions will only serve to further blockade any given excitation. On the other hand, in the case of off-diagonal dipole-dipole interactions this will not be the case, even when all matrix elements $V_{ij}$ are positive. This can be understood by considering two atoms whose dipole-dipole interaction has a strength of $V$. Because the interactions are off-diagonal, the corresponding eigenvalues are $\pm V$. As a result, if two $p$ atoms are each blockading an atom in the ground state, then it becomes possible for the blockade effects to interfere and effectively cancel each other out, allowing the ground state atom to be excited. In a many-body system, this becomes more complicated, with many different atoms taking part in a given excitation.

In order to observe this effect, we will consider both 1D systems, whose matrix elements $V_{ij}$ are all the same sign, and 3D systems, whose matrix elements $V_{ij}$ may be positive or negative. Additionally, the blockade radius will be defined according to the connected correlations between $s$ and $p$ states,
\begin{equation}
\langle \sigma^{pp}_0 \sigma^{ss}_\mathbf{r} \rangle_c = \langle \sigma^{pp}_0 \sigma^{ss}_\mathbf{r} \rangle - \langle \sigma^{pp}_0 \rangle \langle \sigma^{ss}_\mathbf{r} \rangle.
\end{equation}
These correlations describe how a $p$ atom at the origin affects the likelihood there is an $s$ atom at $\mathbf{r}$. When the strength of the dipole-dipole interaction between two atoms is strong compared to the Rabi frequency, the connected correlation will be negative and approximately constant. A negative connected correlation corresponds to the effect of blockade, as it indicates a decreased likelihood for an $s$ atom to be present near a $p$ atom. It is constant for large interaction strengths because increasing the interaction strength further only serves to move a far off-resonant excitation further away from resonance, so the $s$ state will be strongly blockaded in either case.

Unlike in the case of the 1D system, the 3D system can have small interactions for short distances because of the dipole-dipole interaction's angular dependence. As a result, the concept of a blockade radius is slightly modified, so we will instead consider an effective distance 
\begin{equation}
r_\textrm{eff} = r/|1-3 \cos^2 \theta|^{1/3}.
\end{equation}
Under this definition, sites which do not interact with each other are considered as being infinitely far apart. While in reality these nearby sites will affect each other due to higher-order processes even if they do not interact, this effective distance provides a useful way to understand how the effect of blockading is modified in many-body systems.

In Fig.~\ref{corr}, we plot examples of the connected correlations for a 1D system and a 3D system. As expected, we see that for small distances the connected correlations are negative and approximately constant, with the 3D system showing more fluctuations due to many-body effects and the angular dependence of the interactions. As the distance is increased, these correlations drop off to zero, indicating a lack of any correlation due to negligible interaction strength. In 1D, there is some oscillation in the correlations after $r=7$. This likely arises in a similar manner to the emergence of staggered order in other driven-dissipative Rydberg system, in which the blockading of nearby atoms prevents further atoms from being similarly blockaded \cite{Malossi2014,Lee2011,Honing2013}. There are also some outliers in the 3D correlations, which likely arise via a combination of many-body effects, the use of $r_\textrm{eff}$, and artifacts from the cumulant expansion approximation. Finally, we note that the many-body blockade radius is clearly smaller than the two-body blockade radius in both cases, illustrating the presence of anti-blockade effects.

\begin{figure}
\includegraphics[scale=.3]{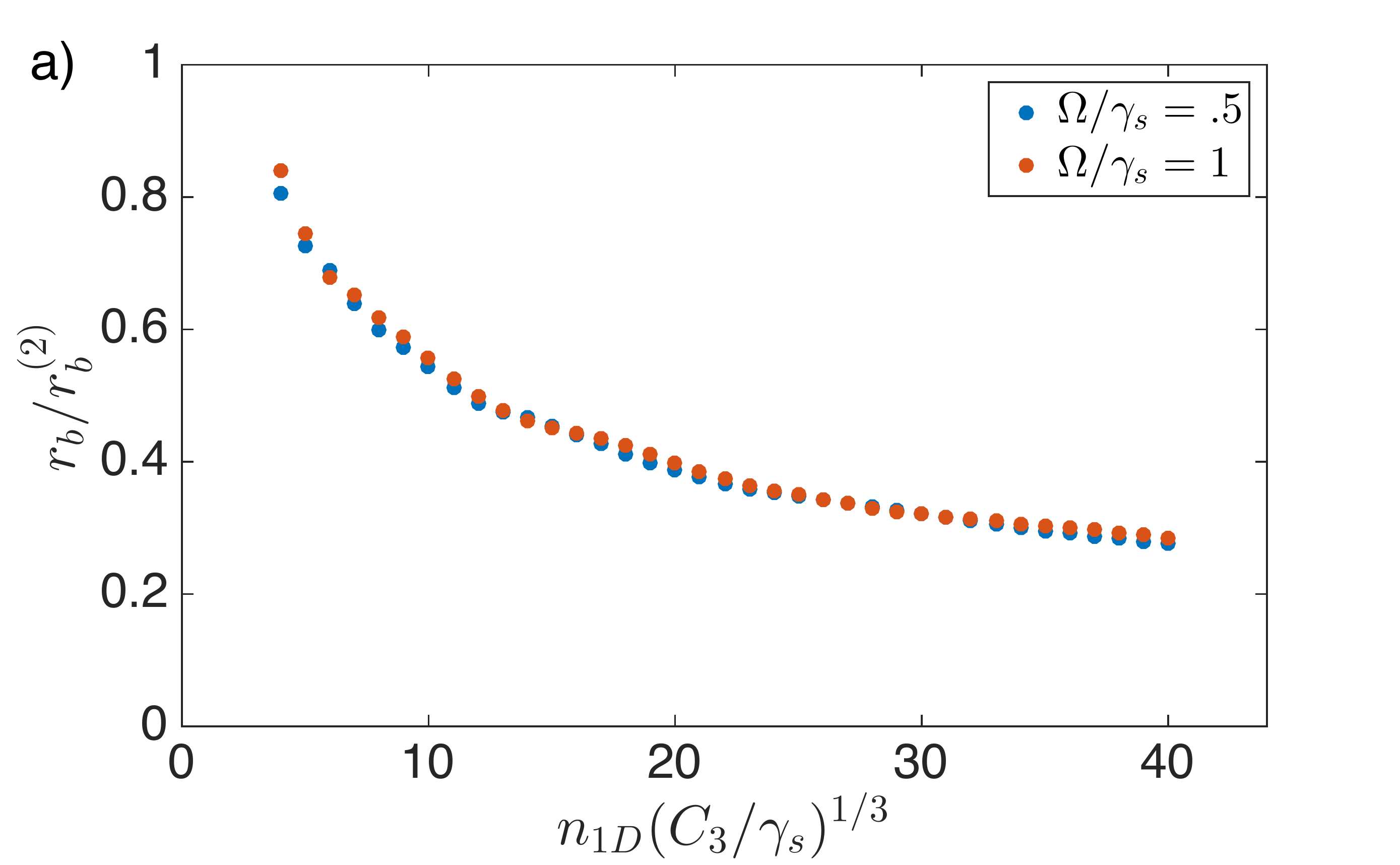}
\includegraphics[scale=.3]{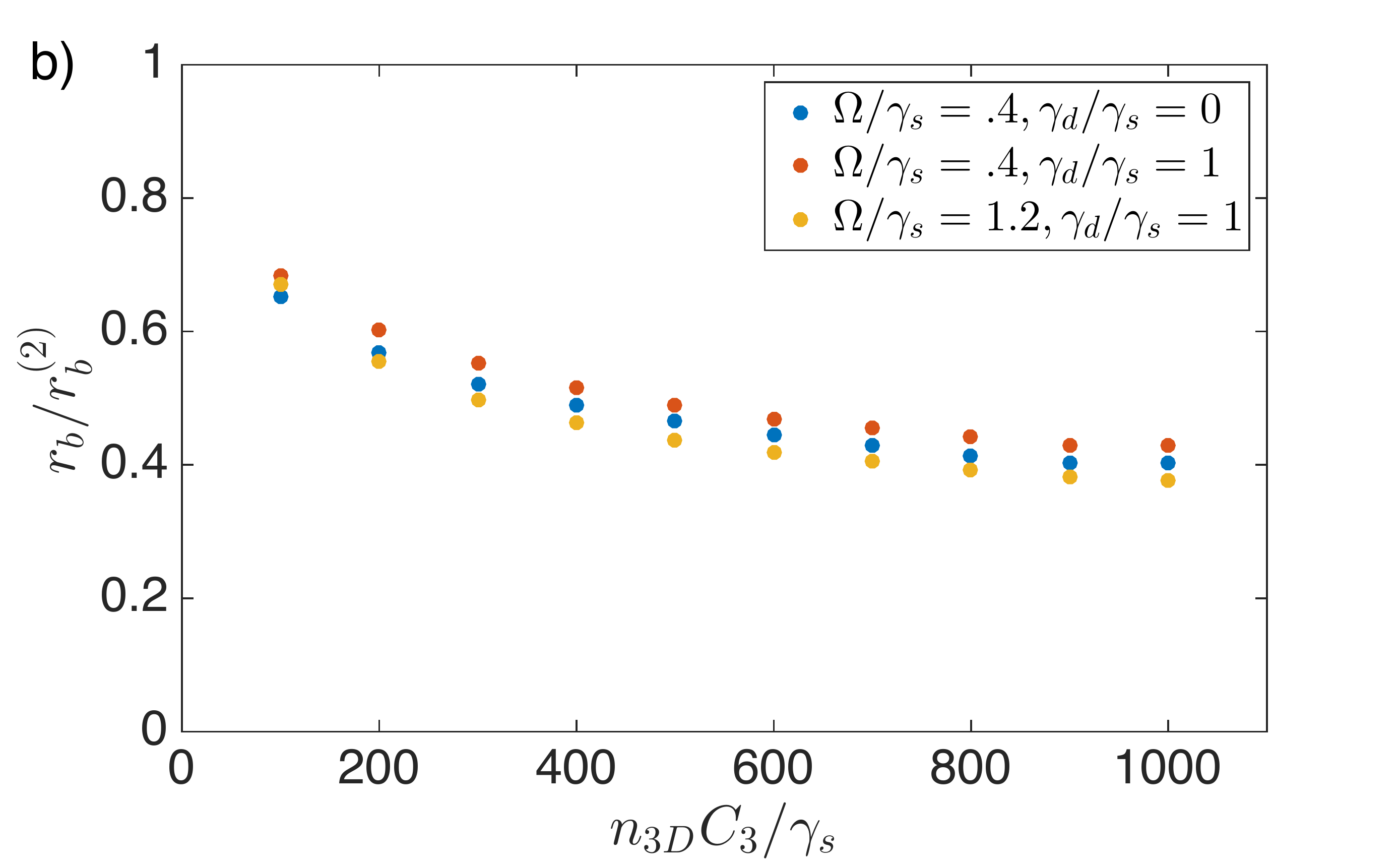}
\caption{Ratio of the many-body blockade radius $r_b$ to the two-body blockade radius $r_b^{(2)}$ as a function of interaction strength. (a) 1D system with $\gamma_d = 0$ for all points. (b) 3D system with examples of both $\gamma_d = 0$ and $\gamma_d \neq 0$. \label{blockrad}}
\end{figure}

In order to extract an effective blockade radius from these connected correlations, we will consider the distance or effective distance at which the correlations decrease by a factor of $1/2$. To reduce the effect of the fluctuations in the 3D system, the correlation at an arbitrary effective distance is defined by an average of the correlations from the cumulant expansion in a range of $\Delta r_\textrm{eff} = 1$, effectively smoothing out the numerics.

In Fig.~\ref{blockrad}, we consider the ratio of the many-body blockade radius to the two-body radius for a variety of parameters. At small densities and interaction strengths, this ratio approaches one, as is expected. However, once we consider larger densities and interaction strengths, the ratio begins to decrease, demonstrating the effect of the competition between blockade and anti-blockade effects. Remarkably, the trend is qualitatively similar for both small and large Rabi frequency, regardless of whether the system is in 1D or 3D. Furthermore, including decoherence does not drastically change the quantitative behavior in 3D. However, this similarity in behavior does not appear to hold for arbitrary $\Omega$. By solving the cumulant expansion equations of motion perturbatively in $\Omega$, it can be shown that to lowest order in $\Omega$, the ratio of the many-body blockade radius to the two-body blockade radius is one. This result reflects the fact that when one goes to sufficiently small Rabi frequencies, the Rydberg population becomes small, so it is rare to have two or more nearby $p$ atoms to give rise to many-body effects.

\subsection{Rydberg Population Scaling}

\begin{figure}
\includegraphics[scale=.3]{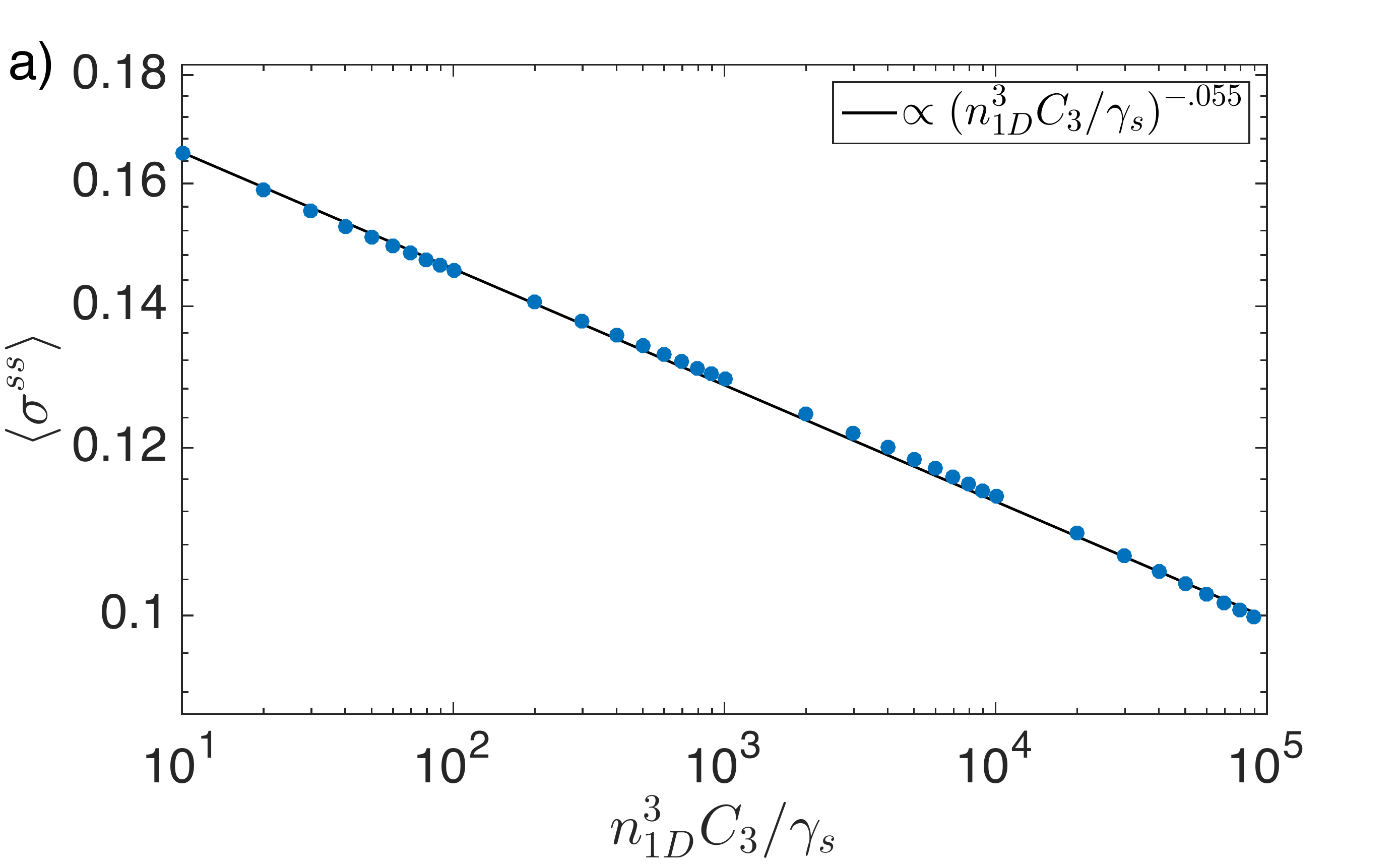}
\includegraphics[scale=.3]{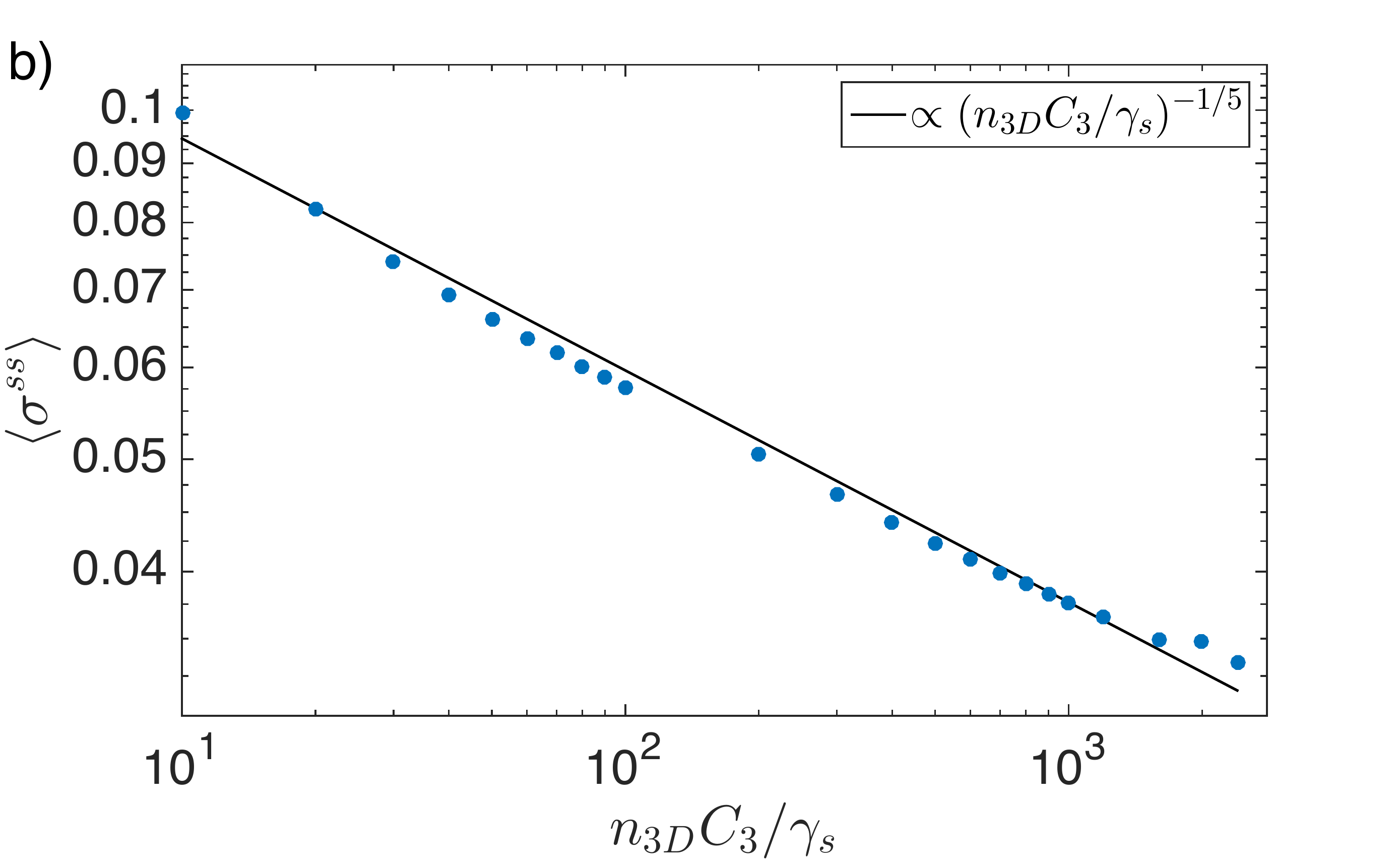}
\caption{Steady-state $s$ state population dependence on interaction strength for $\Omega/\gamma_s = .4$. (a) 1D system with best fit power-law with exponent of $-.055$. (b) 3D system with best fit power-law exponent of $-1/5$. \label{C3scaling}}
\end{figure}

Next, we are interested in understanding how the Rydberg population is affected by dipole-dipole interactions. Although the many-body blockade radius is smaller than the two-body blockade radius at large interaction strengths, both increase as the interaction strength is increased, so we should expect to see a corresponding decrease in the Rydberg population. Fig.~\ref{C3scaling} illustrates the steady-state population's dependence on interaction strength for both 1D and 3D systems. The population appears to decrease according to a power law with a fitted exponent of $-.055$ for 1D and $-1/5$ for 3D, observed over four and two orders of magnitude respectively. 

Particularly in 3D, there is some deviation from purely power law behavior, with a faster fall-off at small interaction strengths compared to large interaction strengths. This is the opposite of what one might normally expect for power law behavior. When there are no interactions, the population is given by some constant, so we would expect a slower fall-off at small interaction strengths. Since the cumulant expansion is more accurate for weak interactions, the fact that we do not see this indicates that if we were to solve the full master equation, we would probably see a faster fall-off at large interactions than what we see here. As a result, we expect the full master equation to result in a scaling behavior much closer to the experimentally observed exponent of $-1/2$ \cite{Goldschmidt2016}. A likely source of this behavior is that at higher interaction strengths, higher-order correlations become more important, and ignoring these correlations ignores relevant blockading effects. However, in order to confirm this hypothesis theoretically, it is important to account for higher-order correlations, which is difficult to achieve in practice.

We are also interested in understanding the population's dependence on the Rabi frequency, which was originally observed to be closer to linear dependence rather than the quadratic behavior of a non-interacting system \cite{Goldschmidt2016}. However, at sufficiently small Rabi frequencies, the density of $s$ excitations will be so small that interactions will become irrelevant, at which point quadratic behavior should be restored. This can be seen by treating the system perturbatively in $\Omega$, which results in $\langle \sigma^{ss} \rangle \approx \frac{4 \Omega^2}{(\gamma_s+\gamma_R)^2}$ to lowest order. This is the same perturbative result as for a non-interacting system.

\begin{figure}
\includegraphics[scale=.3]{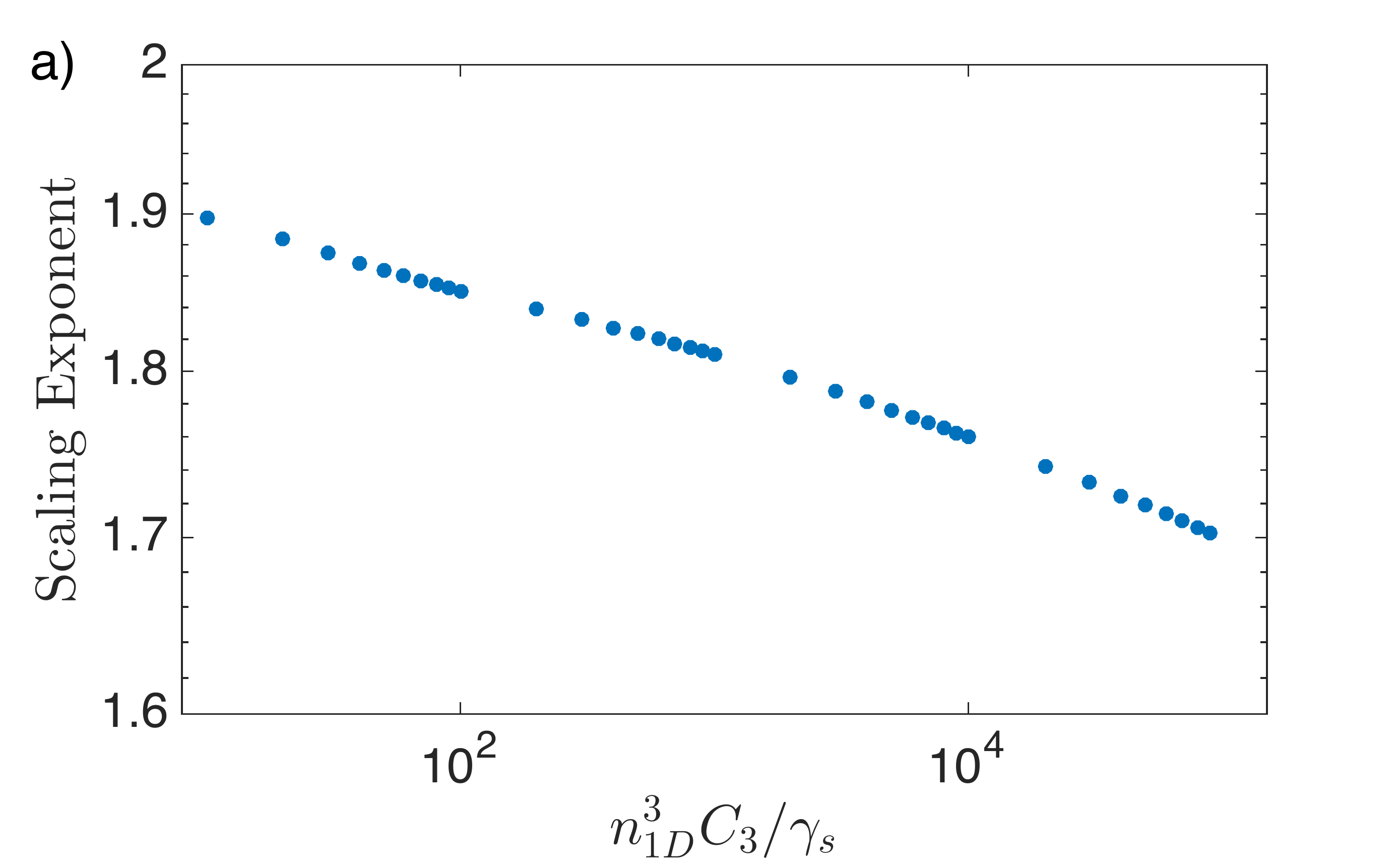}
\includegraphics[scale=.3]{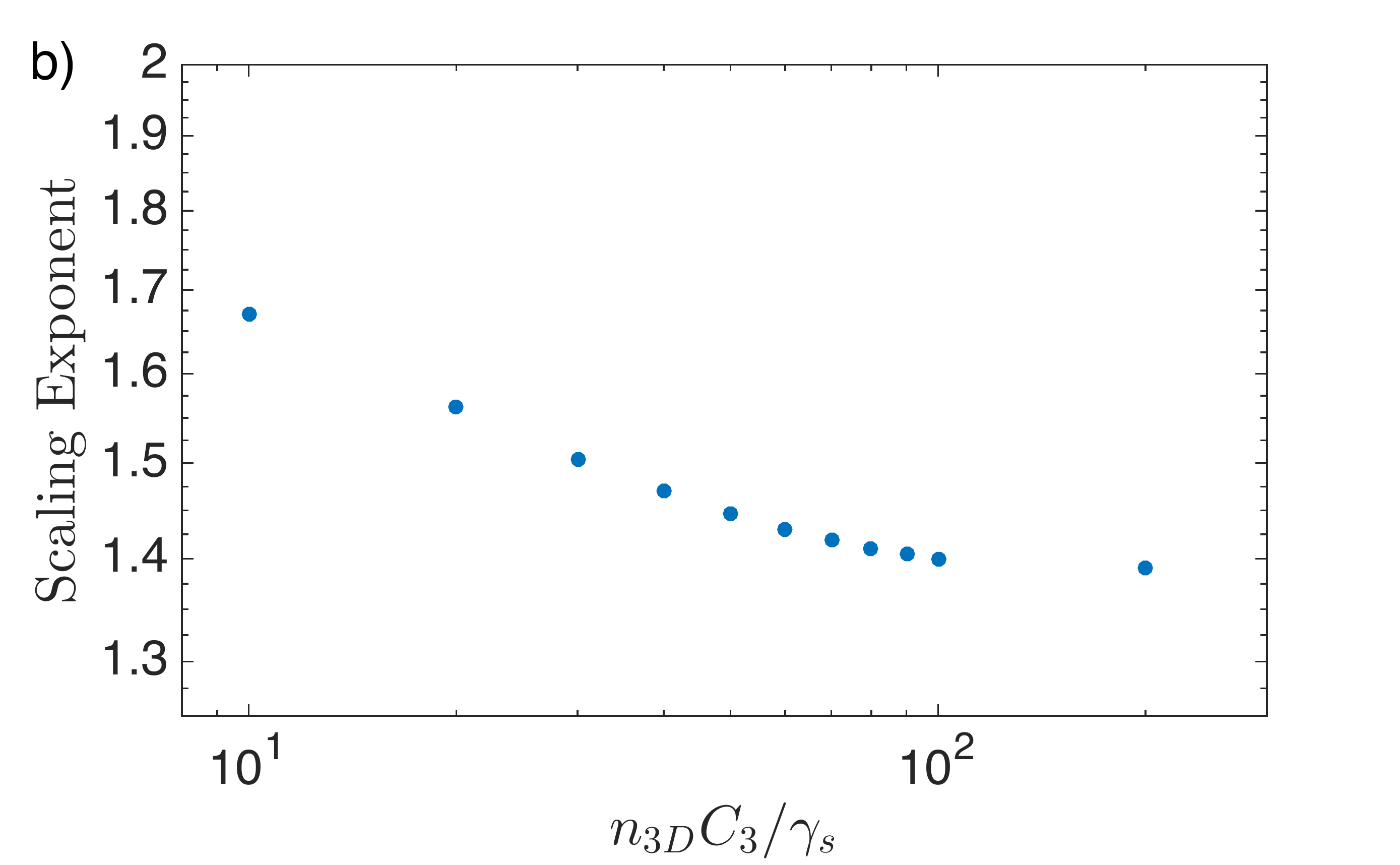}
\caption{Scaling exponent $b$ from fit of $\langle \sigma^{ss} \rangle = a \Omega^b$ for $\Omega/\gamma_s = .05,.1$ as a function of interaction strength. (a) 1D system. (b) 3D system with data only up to $n_{\textrm{3D}} C_3/\gamma_s = 200$ due to finite size effects. \label{exp}}
\end{figure}

We observe this effect in Fig.~\ref{exp}, where we find the fit of the population for two points $\Omega = .05 \gamma_s, .1 \gamma_s$ using the function $\langle \sigma^{ss} \rangle = a \Omega^b$. The value of $b$ is essentially an approximation of the slope on a log-log plot for $\Omega \approx .05 \gamma_s$. As the Rabi frequency is increased, this slope will decrease, transitioning from quadratic behavior towards linear behavior. As the interaction strength is increased, this exponent decreases, indicating that the departure from quadratic behavior is happening at lower Rabi frequencies, allowing for a possible linear behavior over a large range of Rabi frequencies. Additionally, since the Rydberg population is suppressed more for larger interactions, reaching saturation will require stronger Rabi frequencies, expanding the possible range of linear behavior even further. For 3D, we only consider a maximum interaction strength of $n_{\textrm{3D}} C_3/\gamma_s = 200$. This is because past this point, the interactions at the furthest distances we allow become comparable to the small Rabi frequencies considered and the numerics become less accurate. In spite of this restriction on the range of interaction strengths we can consider, we see that the extracted exponent decreases at a much faster rate in the higher dimensional system.

\begin{figure}
\includegraphics[scale=.3]{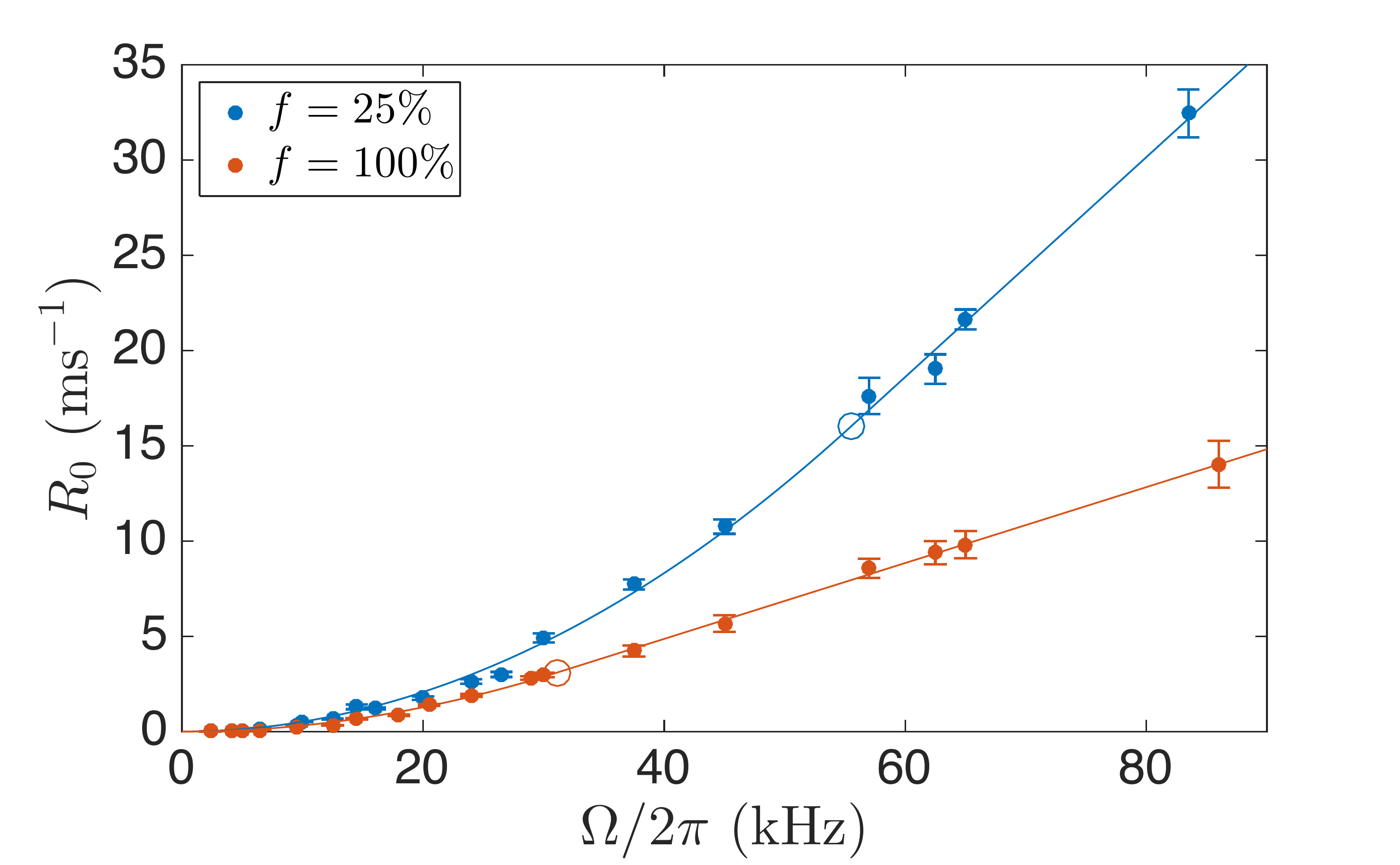}
\caption{Resonant pumping rate as a function of Rabi frequency for two atomic densities, where $f$ corresponds to the fractional density of atoms initially in the driven ground state and $f=100\%$ corresponds to a density of $57$ $\mu$m$^{-3}$. Dots and error bars are from experimental data and lines are from fits of Eq.~(\ref{quadlin}). The empty circles correspond to the fitted crossover Rabi frequency $\Omega_c$, denoting the crossover from quadratic to linear scaling in Rabi frequency. \label{exprabi}}
\end{figure}

In order to determine whether this behavior corresponds to a real effect or simply an artifact of the cumulant expansion approximation, we study this change in scaling behavior experimentally. Using the same experimental setup as in Ref. \cite{Goldschmidt2016}, we consider the scaling behavior for two different densities which differ by a factor of four. In Fig.~\ref{exprabi}, we plot the resonant pumping rate as a function of Rabi frequency for the two different densities. The pumping rate gives the rate at which atoms are pumped out of the relevant three-level system of Fig.~\ref{levels} once a quasi-steady-state has been reached. Further experimental details can be found in Appendix \ref{expmethods}. This pumping rate provides a good approximation of the steady-state population of Rydberg atoms $\langle \sigma^{ss} \rangle \approx \frac{R_0}{\gamma'}$, where $\gamma'$ is the total decay rate from the $s$ state, including decay which takes an atom to an undriven ground state. For both densities, we have determined a best fit using the function
\begin{equation}
R_0 = \left\{ \begin{array}{ll}
a \Omega^2 & \Omega< \Omega_c \\
a \Omega_c (2 \Omega- \Omega_c) & \Omega \geq \Omega_c
\end{array}
\right.
\label{quadlin}
\end{equation}
This describes a continuous, smooth function which changes from quadratic scaling to linear scaling at a critical value of Rabi frequency $\Omega_c$. While in reality the change in scaling behavior may be more gradual, this gives a useful way of determining where the scaling behavior change occurs. We find that $\frac{\Omega_c}{2 \pi} = 31 \pm 1$ MHz for the higher density and $\frac{\Omega_c}{2 \pi} = 55 \pm 2$ MHz for the lower density, clearly illustrating that the scaling behavior changes at smaller Rabi frequencies for higher density samples. We further note that although the quadratic regime is visible at low Rabi frequencies, the corresponding populations are still well below the single-particle limit. This indicates that the quadratic behavior extends beyond the single-particle physics considered above. While this crossover is fairly clear in the experiment, theoretically we see a much more gradual crossover. This could be due to finite size effects or van der Waals interactions, which we have ignored in our model.

\section{Rate Equation Results}

\label{ratesec}

In this section, we will discuss the results of our phenomenological rate equation approach in Eqs. (\ref{rate1}-\ref{rate3}) with the aim of comparing the lineshapes, scaling behavior of the resonant Rydberg population, and the scaling behavior of the linewidths to the experimental results in Ref. \cite{Goldschmidt2016}. Due to computational constraints, we will restrict ourselves to considering 1080 atoms independently and identically distributed according to a 3D Gaussian probability distribution with relative spatial dimensions of $2 \times 4 \times 5$, which is similar to the experimental setup. The density $n_{\textrm{3D}}$ will be taken to be the density at the center of the distribution. Using a uniform probability distribution gives similar results. While the experiment takes place in a lattice, we consider a random distribution to help capture the fact that at any given time, the distribution of the $p$ atoms themselves will be random due to dissipation and will not fully exhibit the structure of the lattice. Although we will vary $n_{\textrm{3D}} C_3$ rather than the total number of atoms, both approaches result in quantitatively similar behavior. For convenience, these rate equations are displayed again below.

\begin{subequations}
\begin{equation}
\dot{s_i} = R_i(g_i - s_i) - (\gamma_s + \gamma_R) s_i,
\end{equation}
\begin{equation}
\dot{p_i} =  \gamma_R s_i - \gamma_p p_i,
\end{equation}
\begin{equation}
\dot{g_i} = -R_i(g_i - s_i) + \gamma_s s_i + \gamma_p p_i,
\end{equation}
\end{subequations}
where the pumping rate $R_i$ is given by
\begin{equation}
R_i = \frac{\Omega^2}{\delta^2 + \Gamma_i^2/4}\Gamma_i,
\end{equation}
and the dephasing rate $\Gamma_i$ is given by
\begin{equation}
\Gamma_i = \gamma_s + \gamma_R + C_3 \left|\sum_{j \neq i} V_{ij} p_j\right|.
\end{equation}
The variables $s_i, p_i, g_i$ refer respectively to the $s, p, g$ populations at site $i$.

The existence of a steady state can be influenced by the manner in which interactions are included in the dephasing. For the choice we are considering here, there are some regions in which no steady-state solution exists. At the edges of such regions, the long-time behavior is periodic, exhibiting limit cycles. Further into these parameter regimes, this periodic behavior likely continues, although the time to reach the limit cycles becomes prohibitive due to the number of atoms in the system. However, in either case the average population of all atoms approaches an approximate steady-state value relatively quickly, with only small deviations from this value as a function of time. Thus we may take a time average of the $s$ state ensemble average in order to find a good approximation of the $s$ population. We find that our phenomenological rate equations produce scaling behavior which is remarkably similar to the experimentally observed scaling behavior as well as very Lorentzian lineshapes.

\begin{figure}
\includegraphics[scale=.3]{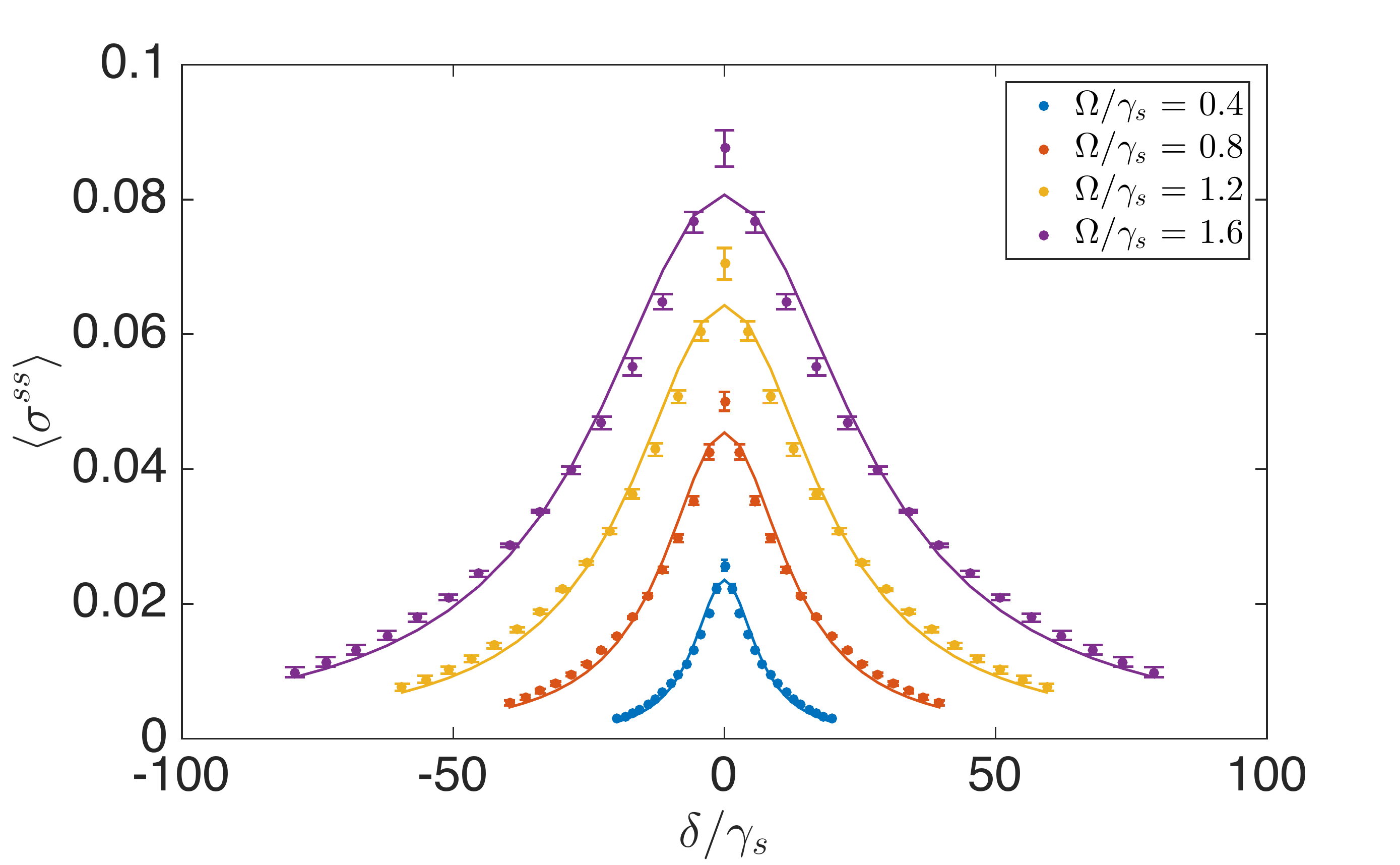}
\caption{Examples of near-Lorentzian lineshapes from inhomogeneous rate equations at $n_{\textrm{3D}} C_3/\gamma_s = 5000$ for several Rabi frequencies.~Error bars indicate standard deviations from five random distributions of atoms and the lines are best-fit Lorentzians. \label{rateshapes}}
\end{figure}

As mentioned before, the exact manner in which the interactions are included in the decoherence can have an effect on the behavior of the steady-state populations. For example, if a homogeneous set of rate equations is used in which the decoherence is merely proportional to $n_{\textrm{3D}} C_3$ times the average $p$ population, this will give reasonable scaling behavior, but it will also result in dome-shaped lineshapes which drop off much faster than a Lorentzian. However, this homogeneous approach ignores the importance of the spatial distribution of the $p$ atoms, which influences the strength and nature of the interactions and thus the decoherence. In order to capture this behavior, some form of inhomogeneity should be included in the rate equations. Our choice of decoherence and atomic distribution provides a simple way of capturing these features and results in more accurate lineshapes.

In Fig.~\ref{rateshapes}, we plot the resulting lineshapes for several different Rabi frequencies. We find that aside from the sharper behavior near resonance, the lineshapes appear to be quite Lorentzian, even at very large linewidths. Another simple choice of decoherence we might make is $\Gamma_i = \gamma_s+\gamma_R + C_3 \sum_{j \neq i} |V_{ij}| p_j$, which only allows decoherence from different sites to add constructively. The resulting lineshapes from this choice would be similar, but they would be more Lorentzian near resonance and drop off faster than a Lorentzian in the wings. Additionally, a steady state is present in all parameter regimes, in contrast to our choice of decoherence.

\begin{figure*}
\includegraphics[scale=.3]{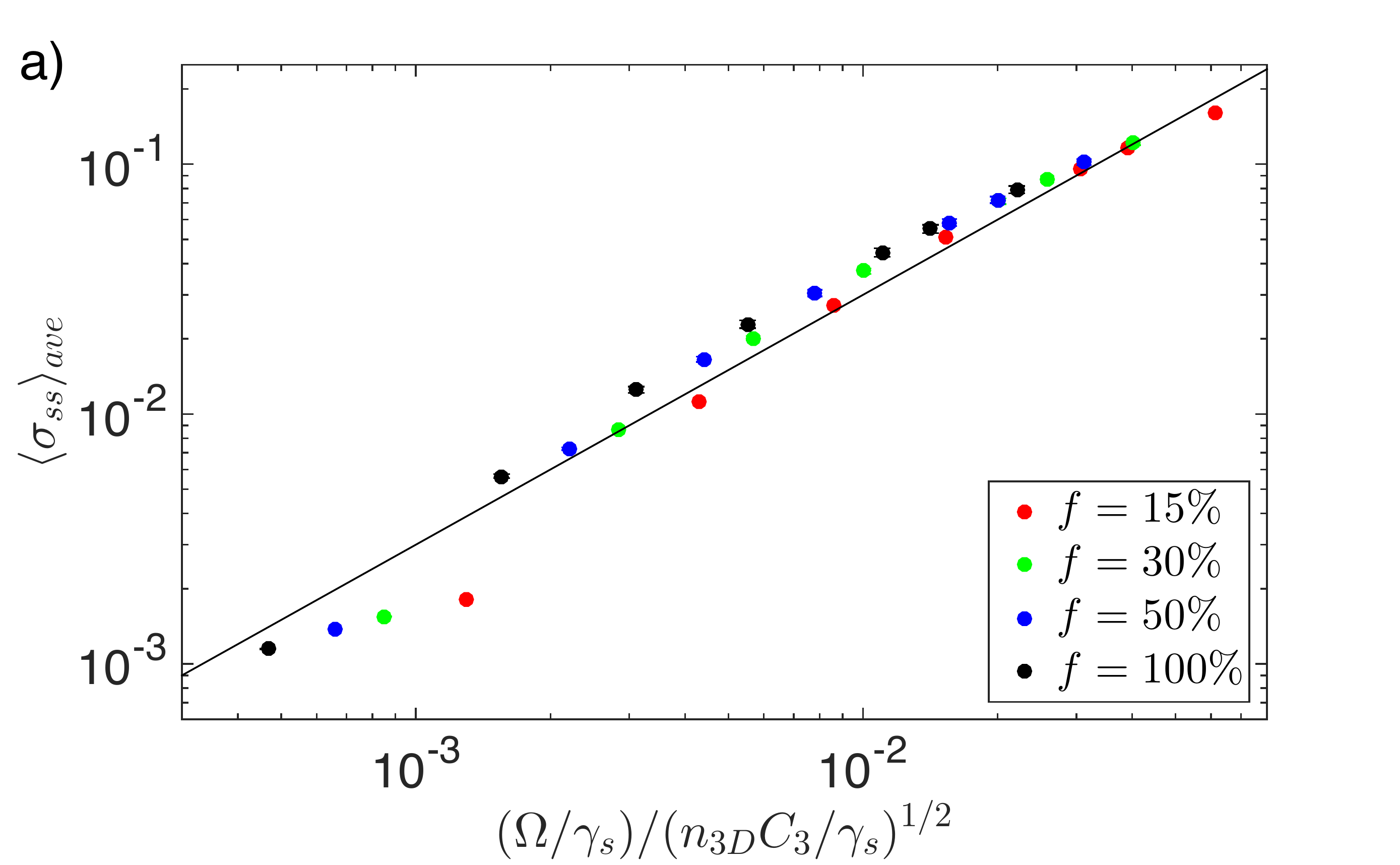}
\includegraphics[scale=.3]{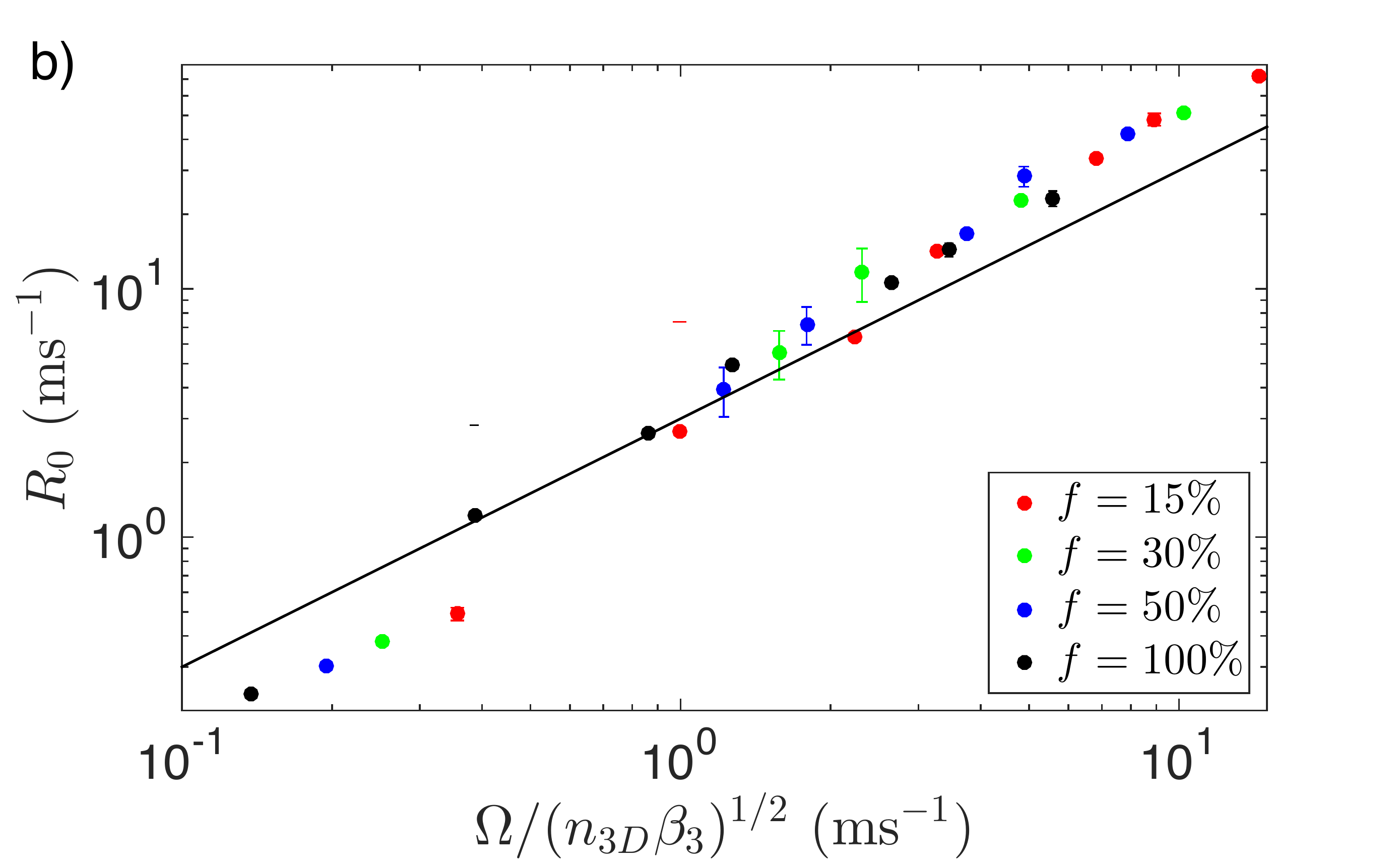}
\caption{Resonant steady-state $s$ population scaling as a function of Rabi frequency and interaction strength. Different points of the same color correspond to different Rabi frequencies and error bars represent fitting uncertainty in the Lorentzian fits. (a) Theoretical rate equation results, where $f = 100\%$ corresponds to $n_{\textrm{3D}} C_3/\gamma_s = 5000$. The solid line is a linear fit with a slope of 3. (b) Experimental resonant pumping rate results from Ref. \cite{Goldschmidt2016} where $f = 100\%$ corresponds to $n_{\textrm{3D}} \beta_3 = 6612$. The solid line is a linear fit with a slope of 3. Note that our definition of $\Omega$ differs from the reference by a factor of two.\label{rescomp}}
\end{figure*}

\begin{figure*}
\includegraphics[scale=.3]{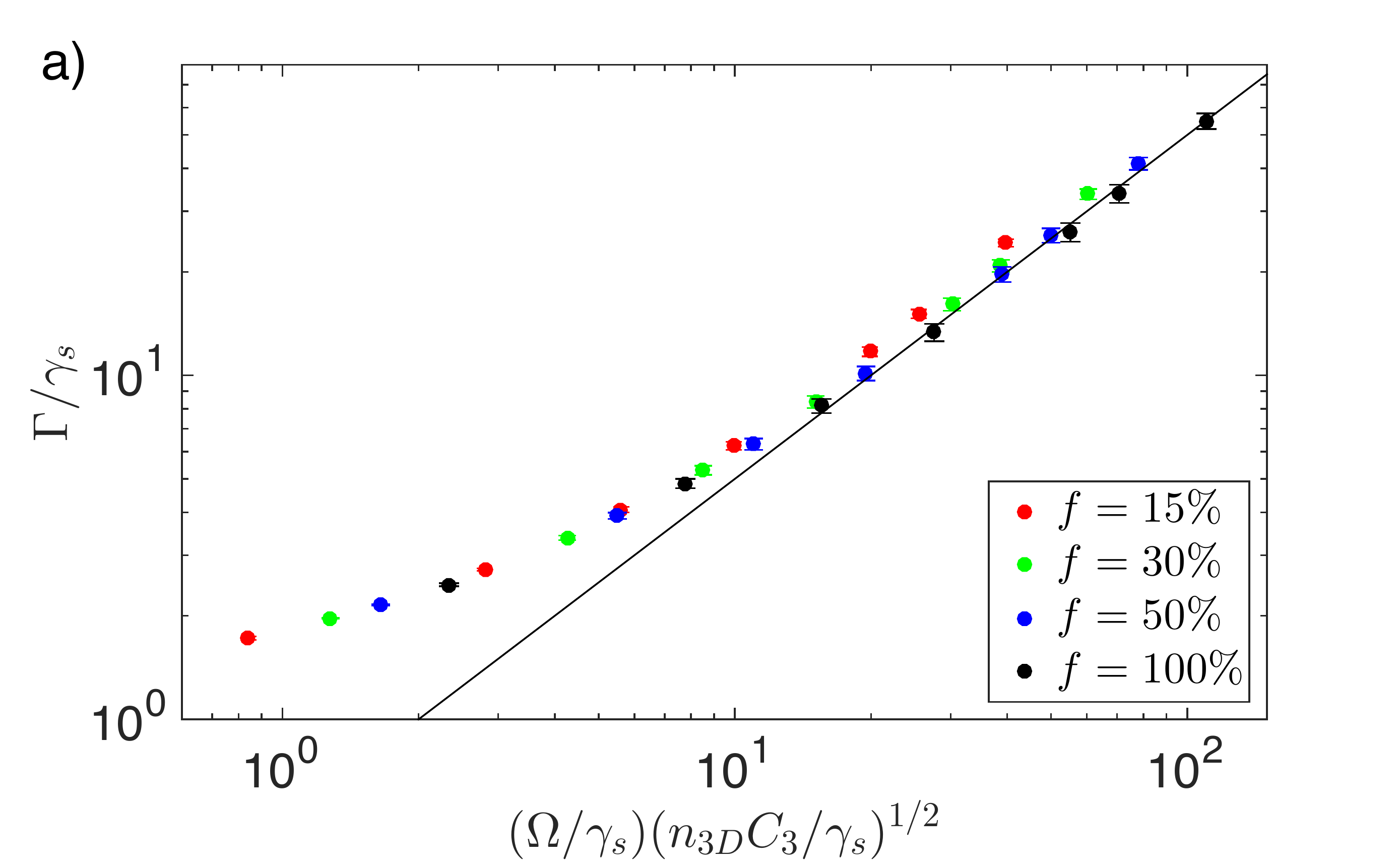}
\includegraphics[scale=.3]{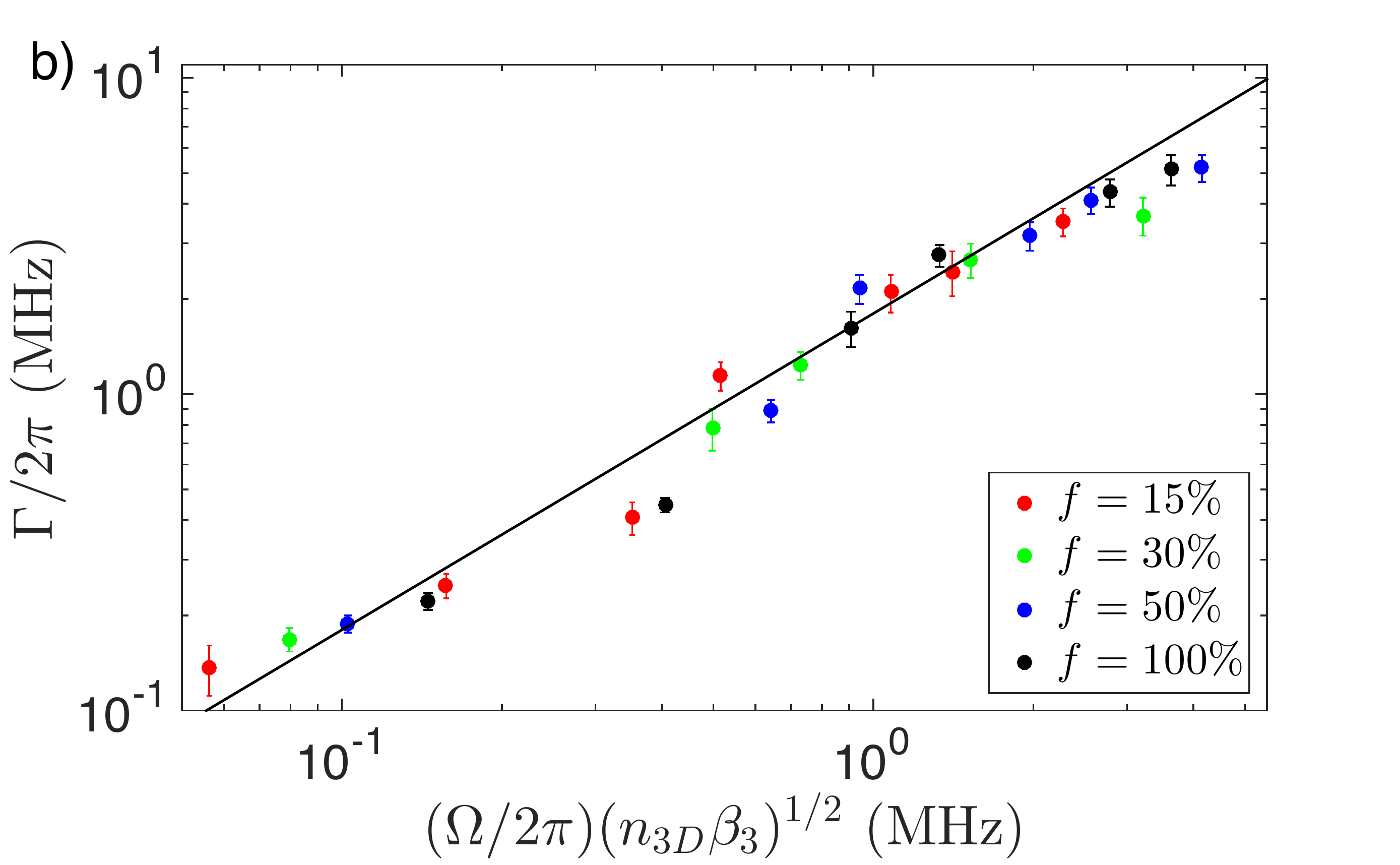}
\caption{Steady-state linewidth scaling as a function of Rabi frequency and interaction strength. Different points of the same color correspond to different Rabi frequencies and error bars represent fitting uncertainty in the Lorentzian fits. (a) Theoretical rate equation results, where $f = 100\%$ corresponds to $n_{\textrm{3D}} C_3/\gamma_s = 5000$. The solid line is a linear fit with a slope of .5. This fit only uses the data from large Rabi frequencies where the bare linewidth does not affect the scaling behavior. (b) Experimental linewidth results from Ref. \cite{Goldschmidt2016} where $f = 100\%$ corresponds to $n_{\textrm{3D}} \beta_3 = 6612$. The solid line is a linear fit with a slope of 1.8. Note that our definition of $\Omega$ differs from the reference by a factor of two. \label{widcomp}}
\end{figure*}

In Fig.~\ref{rescomp}, we compare the resonant population scaling behavior of the rate equation model to the scaling behavior observed experimentally. As before, we can relate the steady-state Rydberg population to the pumping rate via $\langle \sigma^{ss} \rangle \approx \frac{R_0}{\gamma'}$, where $\gamma'$ corresponds to the total decay rate from the $s$ state, including decay which takes the atom to an undriven ground state. Note that $\beta_3 = \sum |C_3^{(np)}| b_{np}/\Gamma_{np}$, where $C_3^{(np)}$ is the corresponding value of $C_3$ for a $p$ state times the root-mean-squared average of the angular dependence, $b_{np}$ are the branching ratios from the driven $s$ state to various $p$ states, and $\Gamma_{np}$ are their corresponding decay rates. This means that $n_{\textrm{3D}} \beta_3$ will be comparable to $n_{\textrm{3D}} C_3/\gamma_s$. Because $\gamma'$ is the same order of magnitude as $\gamma_s$, we should expect similar dependence on $\Omega$ and $n_{\textrm{3D}} C_3$ or $n_{\textrm{3D}} \beta_3$ up to some constant factor. This is in fact the case, with the constant coefficient differing by less than a factor of two. Additionally, the change in behavior between smaller Rabi frequencies and larger Rabi frequencies is quantitatively similar as well, with both exhibiting a slight jump. 

In Fig.~\ref{widcomp}, we compare the linewidth scaling behavior of our model to the scaling behavior observed experimentally. At the lowest Rabi frequencies, the linewidth approaches $\gamma_s$, which is the minimum width possible. Above these lower Rabi frequencies, we find that the general scaling behavior is again the same for theory and experiment, differing only by a constant factor, which in this case is approximately four. While this is not as consistent as for the resonant scaling behavior, it is remarkably consistent considering the simplicity of our model. 

Furthermore, although this is not shown, we have also studied how the scaling coefficients changes as $\gamma_p$ and $\gamma_R$ are varied between $.4 \gamma_s$ and $2 \gamma_s$, which is comparable to the range possible for $^{87}$Rb at $T = 300$ K. We find that the scaling coefficient for the resonant population and linewidth varies approximately according to $\sqrt{\gamma_p / \gamma_R}$ and $\sqrt{\gamma_R / \gamma_p}$ respectively, which is consistent with the definition of $\beta_3$. This is natural since $\gamma_p/\gamma_R$ corresponds to the ratio of $s$ atoms to $p$ atoms, so a higher ratio results in stronger dephasing in the same way that an increase in the interaction strength results in stronger dephasing. However, if we were to extend the range of possible $\gamma_p$ and $\gamma_R$ much further, this behavior will start to break down.

While the general scaling behavior of Ref. \cite{Goldschmidt2016} is captured very well, there are two areas in which the rate equations fail qualitatively. The first qualitative failure is in the transition from quadratic to linear scaling in Rabi frequency. This model predicts the resonant linear behavior to continue into much smaller Rabi frequencies than observed experimentally in Fig.~\ref{exprabi} (theory not shown). A possible reason for this is that at low Rabi frequencies, there is a small number of Rydberg atoms, so that the exact details of their interactions and correlations become more important and cannot be treated simply as dephasing. Another possible reason is that we have neglected van der Waals interactions, which may be more important in this regime. The second qualitative failure is the time required to reach steady state, which the model predicts to be much longer than observed experimentally, as noted for the homogeneous rate equations in Ref. \cite{Boulier2017}. This is most likely because the exact details of the interactions and correlations are more important when there is a small number of Rydberg atoms. Rather than a diffuse population of $p$ atoms slowly increasing the excitation rate, there is initially a single $p$ atom which immediately brings directly into resonance many other possible excitations, leading to highly-correlated growth dynamics.

\section{Conclusion and Outlook}

We have investigated the effect that dissipation induced dipole-dipole interactions have in a driven-dissipative Rydberg system using a cumulant expansion approach and phenomenological inhomogeneous rate equations. For the cumulant expansion approach, we showed that a modified many-body blockade radius picture arises, leading to steady-state populations which scale with the interaction strength like a power law. Additionally, we demonstrated a trend away from quadratic scaling in Rabi frequency at low Rabi frequencies for strong interactions. We theoretically predicted and experimentally observed that this transition occurs earliest for high densities. While the cumulant expansion behaves qualitatively similar to experimental observations, it is insufficient for quantitative agreement. This is because in spite of the large amount of dissipation, the strong, long-range nature of the dipole-dipole interaction gives rise to important many-body correlations which need to be taken into account. However, with a simple choice of phenomenological inhomogeneous rate equations in which decoherence is proportional to the interaction strength of nearby $p$ atoms, we found remarkable quantitative agreement with the experimental results of Ref. \cite{Goldschmidt2016}, although the rate equations fail to properly capture low Rabi frequency behavior and early time dynamics, where the actual structure of the correlations is particularly important. 

In order to fully understand the underlying physics which gives rise to the anomalous Rydberg broadening, further theoretical and experimental study is necessary. While we have gone beyond mean field theory by including second-order connected correlations, there are other possible routes as well, such as Keldysh field theory \cite{Maghrebi2016,Sieberer2016} or cluster mean-field approaches \cite{Jin2016}. If one can determine which high-order correlations are likely to be important with a reasonable degree of accuracy, this could provide a better way to reduce the exponential number of equations of motion while still capturing the effects of high-order correlations. The success of our phenomenological rate equations may also provide insight into other systems involving dipole-dipole interactions or a path towards a more rigorous derivation of similar rate equations, as has been done for the case of diagonal interactions in Ref. \cite{Lesanovsky2013}. Furthermore, the regimes where the rate equations performed poorly were where the Rydberg population was smaller, so they may be amenable to methods which take advantage of this. This regime is also where there is likely to be an intseresting interplay between dipole-dipole interactions and van der Waals interactions, which we have neglected here. So far, both theory and experiment have been primarily focused on the effect of the interactions on the total Rydberg population, so determining the details of the many-body correlations theoretically and experimentally remains an interesting open problem.

\begin{acknowledgments}
We thank Ana Maria Rey, Bihui Zhu, Michael Fleischhauer, Susanne Yelin, Zhe-Xuan Gong, Anzi Hu, and Michael Foss-Feig for helpful discussions. The authors acknowledge funding from ARL CDQI, NSF PFC at JQI, ARO, AFOSR, ARO MURI, and NSF QIS. T.B. acknowledges the support of the European Marie Sklodowska-Curie Actions (H2020-MSCA-IF-2015 Grant 701034). RMW acknowledges partial support from the National Science Foundation under Grant No.~PHY-1516421.
\end{acknowledgments}

\appendix

\section{Gutzwiller Mean Field Theory}
\label{MFT}

In Sec.~\ref{modelsec}, we motivated our use of a cumulant expansion approximation due to the fact that Gutzwiller mean field theory fails to provide any insight into our model. In this Appendix, we provide the reasons for this failure. Using an inhomogeneous Gutzwiller mean-field approximation, we assume the density matrix has the form
\begin{equation}
\rho = \bigotimes_i \rho_i,
\end{equation}
which assumes there are no correlations between different atoms \cite{Rokhsar1991, Diehl2010}. The method is inhomogeneous in the sense that each atom has its own density matrix, whereas in homogeneous Gutzwiller mean field theory all atoms have the same density matrix. This results in an effective local Hamiltonian

\begin{equation}
H_i^{\textrm{eff}} = -\delta \sigma_i^{ss} + \Omega (\sigma_i^{sg} + \sigma_i^{gs}) + \sum_j [V_{ij} \sigma_i^{ps} \langle \sigma_j^{sp} \rangle + H.c.].
\end{equation}

Under this approximation, the interactions behave as an effective driving term between the $s$ and $p$ states whose strength and phase are determined by the $\langle \sigma^{sp} \rangle$ coherences of the surrounding atoms. This explicitly assumes a breaking of the $U(1)$ symmetry $|p \rangle \to e^{i \phi} |p \rangle$. If it is not broken, then the system behaves as if there are no interactions. Additionally, in the full master equation's steady state, the ratio of $s$ to $p$ atoms is fixed because the number of $p$ atoms only changes due to single-site decay processes. However, under the mean-field approximation, the effective drive between $s$ and $p$ states will inevitably change this ratio in steady state.

If we want to keep the ratio of $s$ to $p$ atoms reasonably close to the true value, the effective Rabi frequency must be small. In this limit, we can easily solve perturbatively for the steady state value of $\langle \sigma^{sp} \rangle$ as a function of the effective Rabi frequency $\Omega_\textrm{eff} = \sum_j V_{ij} \langle \sigma_j^{sp} \rangle$

\begin{equation}
\langle \sigma_i^{sp} \rangle \approx \frac{i (\langle \sigma^{ss}_{V=0} \rangle - \langle \sigma^{pp}_{V=0} \rangle)}{i \delta - \frac{\gamma_s + \gamma_p + \gamma_R}{2}} \Omega_\textrm{eff},
\end{equation}
where $\langle \sigma^{ss}_{V=0} \rangle$ and $ \langle \sigma^{pp}_{V=0} \rangle$ are the $s$ and $p$ populations with no interactions. In this limit, the coherence depends linearly on the effective Rabi frequency, which can be written as a matrix equation
\begin{equation}
\langle \sigma_i^{sp} \rangle \approx C \sum_j V_{ij} \langle \sigma_j^{sp} \rangle,
\label{mat}
\end{equation}
where $C$ is some complex constant with nonzero imaginary part. Equation (\ref{mat}) may be thought of in terms of finding the eigenvector associated with an eigenvalue $1/C$ of the matrix defined by $V_{ij}$ where $V_{ii} =0$. However, since $V_{ij}$ is a symmetric, real matrix, it has only real eigenvalues, so $1/C$ cannot be an eigenvalue and the only solution to Eq. (\ref{mat}) is $\langle \sigma_i^{sp} \rangle = 0$. Thus the only possibility of a low effective Rabi frequency mean-field solution with nonzero coherences is one which is not constant in time, such as a limit cycle.

In order to determine whether other nontrivial solutions are possible, we initialize a cubic lattice of randomized density matrices for each lattice site and evolve the system according to the mean field equations of motion. This was done for a variety of numerically feasible parameters, while the nearest neighbor interaction strength remained at least two orders of magnitude above $\Omega$ and all decay rates.

In all cases, including those with large initial $\Omega_\textrm{eff}$, we found that the $\langle \sigma^{sp}\rangle$ coherences all decay to zero in steady state, in which case the system behaves as if there are no interactions. This would not occur if the interactions were of the form
\begin{equation}
\sum_{i \neq j} V_{ij} \sigma^{ss}_i \sigma^{pp}_j.
\end{equation}

Collective decay between the $s$ and $p$ states does result in nonzero $\langle \sigma^{sp} \rangle$ coherences in steady state, but we find the effect of interactions when collective decay is included has been found numerically to be small. Furthermore, the experimental results in Refs. \cite{Goldschmidt2016,Boulier2017} indicate that collective decay is not the source of the observed broadening and is suppressed by the dipole-dipole interactions.

\section{Cumulant Expansion Equations of Motion}

In this Appendix, in order to illustrate how the cumulant expansion approximation truncates the hierarchy of differential equations, we will present example derivations for a single-atom expectation value $\langle \sigma^{sg}_i \rangle$ as well as a two-atom expectation value $\langle \sigma_i^{sp} \sigma_j^{pg} \rangle$ while taking advantage of the symmetries mentioned in Sec.~\ref{cumulantsec}. In the full master equation, we can consider the evolution of the expectation value of an operator $\mathcal{O}$ via $\partial_t \langle \mathcal{O}\rangle = \Tr (\dot{\rho} \mathcal{O})$. Thus the corresponding differential equations for the two operators are
\begin{widetext}
\begin{equation}
\partial_t \langle \sigma_i^{sg} \rangle = i \Omega(\langle \sigma_i^{gg} \rangle - \langle \sigma_i^{ss} \rangle) + i \sum_{k \neq i} V_{k i} \langle \sigma_k^{sp} \sigma_i^{pg} \rangle 
- \frac{\gamma_s + \gamma_R + 2 \gamma_d}{2} \langle \sigma_i^{sg} \rangle,
\end{equation}
\begin{equation}
\begin{aligned}
\partial_t \langle \sigma_i^{sp} \sigma_j^{pg} \rangle = & \ i \Omega ( \langle \sigma_i^{gp} \sigma_j^{pg} \rangle - \langle \sigma_i^{sp} \sigma_j^{ps} \rangle) 
- \frac{\gamma_s+\gamma_R+2 \gamma_p + 3\gamma_d}{2} \langle \sigma_i^{sp} \sigma_j^{pg} \rangle
\\
&+ i \sum_{k \neq i} V_{k i} ( \langle \sigma_i^{pp} \sigma_k^{sp} \sigma_j^{pg} \rangle - \langle \sigma_i^{ss} \sigma_k^{sp} \sigma_j^{pg} \rangle)
+ \ i \sum_{k \neq j} V_{k j} \langle \sigma_i^{sp} \sigma_j^{sg} \sigma_k^{ps} \rangle,
\end{aligned}
\end{equation}
\end{widetext}
where the sums are only over $k$. Note that some of the three-atom operators may sometimes be two-atom operators if two of the indices are the same, in which case no approximation is necessary and they are treated exactly.

In the above equations of motion, only the interaction terms couple operators involving a different number of atoms. The driving terms and decay terms always couple to the same sites. Additionally, since the interaction is composed of only two-atom terms, $n$-atom operators can only couple to operators involving $n$ or $n\pm 1$ sites. As a result, assuming three-atom connected correlations to be zero implies all higher-order connected correlations are zero, truncating the hierarchy of equations that results from the interactions.

Once we apply translational invariance, single-atom expectation values are site-independent, e.g. $\langle \sigma_i^{sg} \rangle = \langle \sigma^{sg} \rangle$, and two-atom expectation values depend only on their displacement vector, e.g. $\langle \sigma_i^{sp} \sigma_j^{pg} \rangle = \langle \sigma_0^{sp} \sigma_{j-i}^{pg} \rangle$. Furthermore, the $U(1)$ symmetry of $|p \rangle \to |p \rangle e^{i \phi}$ implies terms like $\langle \sigma_i^{sp} \rangle$ or $\langle \sigma^{pg}_i \sigma^{ss}_j \rangle$ are zero in steady state. Applying the cumulant expansion approximation, the terms in the equations of motion due solely to the interactions become
\begin{widetext}
\begin{equation}
\partial_t \langle \sigma^{sg} \rangle = \cdots + i \sum_{j \neq 0} V_{0 j} \langle \sigma_0^{sp} \sigma_j^{pg} \rangle,
\end{equation}
\begin{equation}
\partial_t \langle \sigma^{sp}_0 \sigma^{pg}_i \rangle = \cdots + i V_{0 i} \langle \sigma^{pp}_0 \sigma^{sg}_i \rangle 
+i \sum_{j \neq 0,i} V_{j i} \langle \sigma_0^{sp} \sigma_j^{pg} \rangle (\langle \sigma^{pp} \rangle - \langle \sigma^{ss} \rangle) 
+ i \sum_{j \neq 0,i} V_{j i} \langle \sigma_0^{sp} \sigma_j^{ps} \rangle \langle \sigma^{sg} \rangle.
\end{equation}
\end{widetext}
Note that there are two types of interaction terms present above. The first involves terms whose interaction strength and two-atom operator correspond to the same atoms, while in the second only one index matches.

\label{cumulant}

\section{Quantum Trajectories}
\label{traj}

In this Appendix, we verify that the cumulant expansion approach is a reasonable approach by comparing it to the exact numerical approach of quantum trajectories \cite{Dalibard1992,Dum1992,Plenio1998,Daley2014}. However, the quantum trajectories approach can be applied for at most 10 atoms due to computational constraints, which puts a limit on the range of $C_3$ we can consider if we want to keep boundary effects to a minimum. Here, we focus on a 1D lattice of atoms with periodic boundary conditions. To take this into account, the interaction between two given atoms is taken to be 
\begin{equation}
V_{ij} = \frac{C_3}{r_1^3}+\frac{C_3}{r_2^3},
\end{equation}
 where $r_{1,2}$ are the two smallest distances between atoms $i$ and $j$. 

We can consider in general a quantum master equation of the following form
\begin{equation}
\dot{\rho} = -i [H,\rho] + \sum_i \gamma_i \left(\mathcal{O}_i \rho \mathcal{O}^\dagger_i - \frac{1}{2}\{\mathcal{O}_i^\dagger \mathcal{O}_i, \rho \} \right),
\end{equation}
which may be rewritten in terms of an effective non-Hermitian Hamiltonian and recycling terms
\begin{subequations}
\begin{equation}
\dot{\rho} = -i (H_{\textrm{eff}} \rho - \rho H_\textrm{eff}^\dagger) + \sum_i \gamma_i \mathcal{O}_i \rho \mathcal{O}_i^\dagger,
\end{equation}
\begin{equation}
H_\textrm{eff} = H - \frac{i}{2} \sum_i \gamma_i \mathcal{O}_i^\dagger \mathcal{O}_i.
\end{equation}
\end{subequations}

Rather than considering the evolution of the density matrix, we will instead consider stochastic evolution of a normalized state $| \psi(t) \rangle$ according to $H_\textrm{eff}$. As a result of the non-Hermitian part of the effective Hamiltonian, the norm of $| \psi(t) \rangle$ is not conserved, and after a time $dt$ it will have a norm of $\langle \tilde{\psi}(t+dt)| \tilde{\psi}(t+dt)\rangle = 1 - p$. The deviation $p$ corresponds directly to the probability that a quantum jump has occurred due to the Lindbladian. In the case where several possible types of quantum jumps are possible, as is the case here, each process is weighted according to 
\begin{subequations}
\begin{equation}
p_i = w_i p,
\end{equation}
\begin{equation}
w_i = \frac{\gamma_i \langle \psi(t)| \mathcal{O}_i^\dagger \mathcal{O}_i | \psi(t) \rangle}{\sum_i \gamma_i \langle \psi(t)| \mathcal{O}_i^\dagger \mathcal{O}_i | \psi(t) \rangle}.
\end{equation}
\end{subequations}

Thus with probability $p_i$ the new state is
\begin{equation}
|\psi(t+dt) \rangle = \mathcal{O}_i | \psi(t) \rangle/\sqrt{\langle \mathcal{O}_i^\dagger \mathcal{O}_i \rangle},
\end{equation} 
and with probability $1-p$ the new state is
\begin{equation}
|\psi(t+dt) \rangle = |\tilde{\psi}(t+dt) \rangle /\sqrt{1-p}.
\end{equation}

\begin{figure}
\includegraphics[scale=.3]{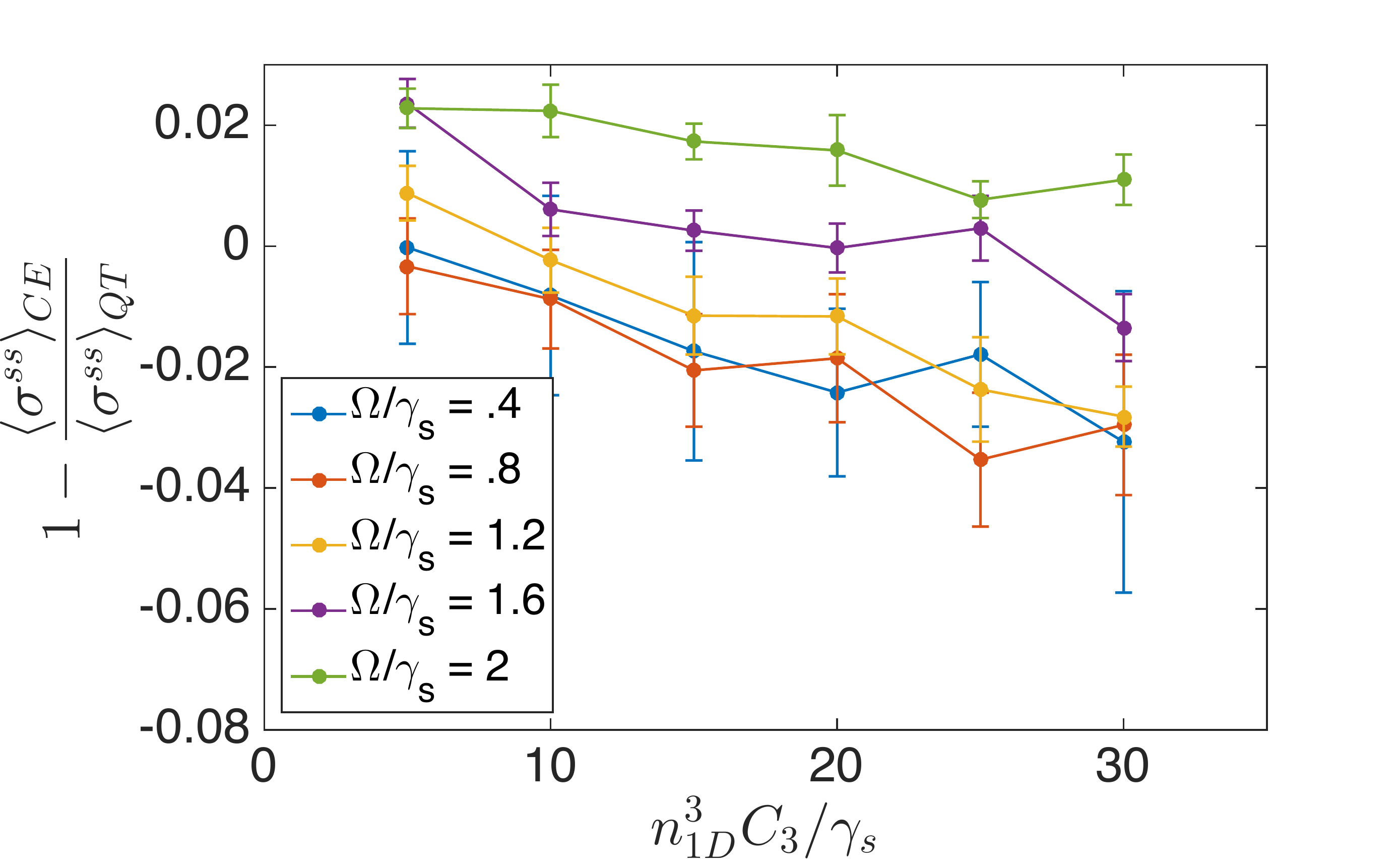}
\caption{Relative error of cumulant expansion (CE) with respect to quantum trajectories (QT). Error bars denote standard error from quantum trajectories sampling. The total sampling time is 4500 $\gamma_s^{-1}$.}
\label{comp}
\end{figure}

In contrast to a density matrix approach, there is no specific steady-state $|\psi\rangle$ which is constant in time. Instead, we extract the corresponding steady-state density matrix by considering time averages of $|\psi\rangle$ once it has evolved sufficiently long to exhibit steady-state behavior. This is effectively equivalent to averaging over many runs to a specific time which is large compared to the steady-state relaxation time. In Fig.~\ref{comp}, we compare the results from quantum trajectories to the results from cumulant expansion. We see that at least in the limit of small interaction strengths and densities, the steady-state error due to the cumulant expansion is not too large.

\section{Experimental Methods}
\label{expmethods}

\begin{figure}[!t]
\includegraphics[scale=.33]{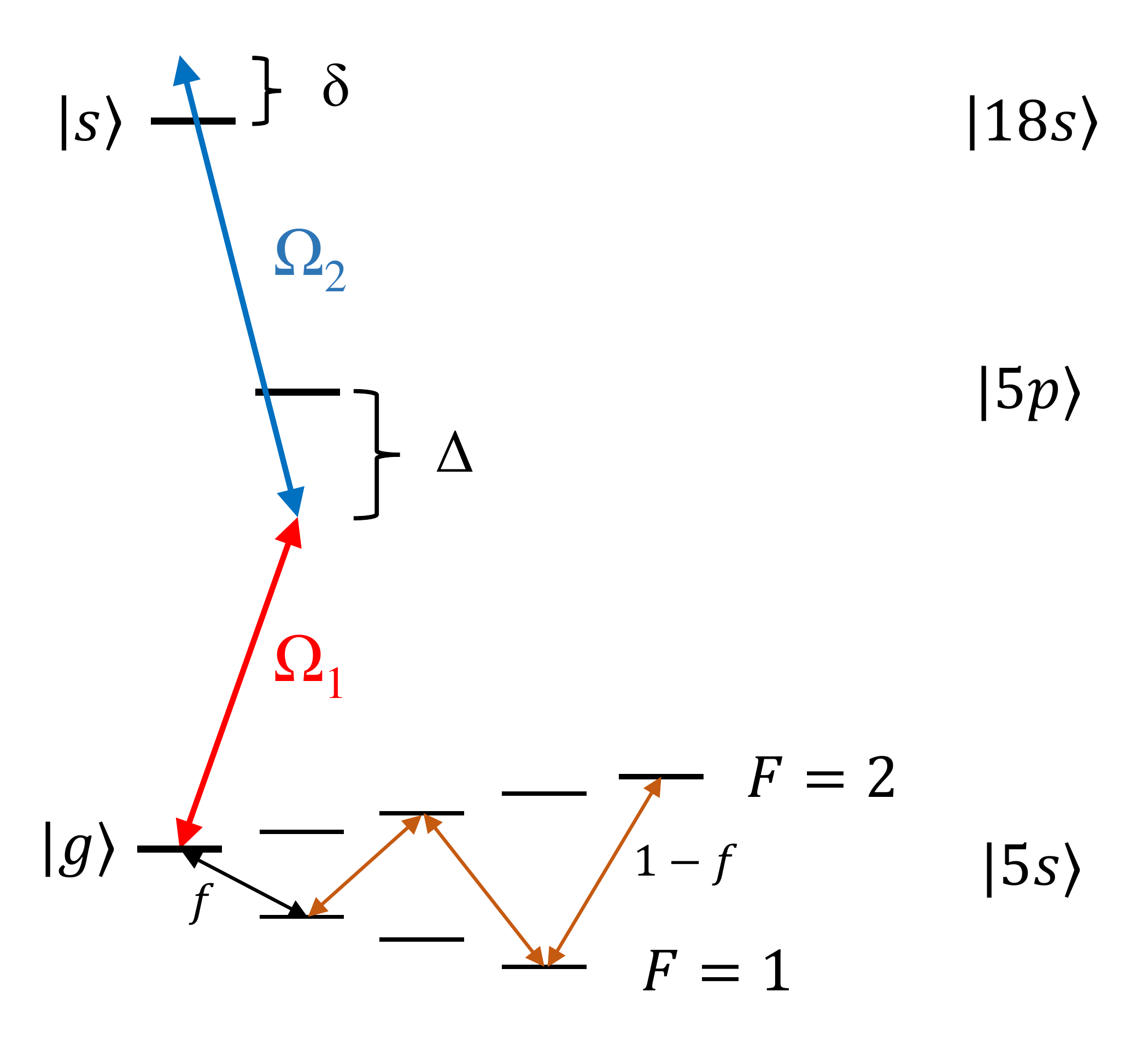}
\caption{Experimental excitation and measurement scheme. The ground-state manifold is initially populated with a fraction $f$ in the $|g \rangle = |F=2, m_F =-2\rangle$ state and $1-f$ in the $|F=2,m_F = 2\rangle$ state using three applications of microwave rapid adiabatic passage. The $|F=2, m_F =-2\rangle$ state is then driven via an off-resonant two-photon transition through the $5p_{1/2}$ state to the $18s_{1/2}$ state with intermediate detuning $\Delta$, two-photon detuning $\delta$, and lower and upper Rabi frequencies $\Omega_1$ and $\Omega_2$. \label{expscheme}}
\end{figure}

In this Appendix, we describe the experimental methods used in the main text. The details of the experimental setup are described in Refs. \cite{Lin2009,Goldschmidt2016}. In a nutshell, the basis for the apparatus is a \Rb  Bose-Einstein condensate (BEC) machine producing a BEC composed of $N\approx 4 \times 10^{4}$ atoms every 16 seconds. We excite the atoms to the $18s_{1/2}$ state using a two-photon transition via the $5p_{1/2}$ state with intermediate detuning $\Delta/2\pi\approx 240$ MHz. The lower and upper Rabi frequencies are independently calibrated to $\Omega_1/2\pi=0$~MHz to $5$~MHz and $\Omega_2/2\pi\approx 12.5$~MHz. In keeping with the notation of the main text, these are both half the typical definition of the Rabi frequency. The two lasers are locked to the same high-finesse optical cavity with $<10$~kHz linewidth and are tuned for the transition $|g \rangle = |5s,F=2,m_F=-2\rangle \rightarrow |s \rangle = |18s,F=2,m_F=-2\rangle$. The BEC is created in the $|F=1, m_F=-1\rangle$ state, and we control the fraction $f$ transferred to $|5s, F=2, m_F=-2\rangle$ via microwave rapid adiabatic passage. The remaining atoms are then transferred to the shelving state $|F=2, m_F=2\rangle$. This offers control over the fractional density of atoms participating to the Rydberg excitation. This process is shown in Fig.~\ref{expscheme}.

The post-excitation populations in the ground hyperfine manifold are separated in time-of-flight with a Stern-Gerlach magnetic field gradient and measured via absorption imaging. Experiments are done in a 3D optical lattice made with 812 nm light, resulting in a lattice spacing of 406 nm. This distance is comparable to the $18s$ van der Waals blockade radius. We measure the resonant Rydberg pumping rate $R_{0}$ as a function of the two-photon Rabi frequency $\Omega = \Omega_1 \Omega_2/\Delta$ for two fractional densities $f$. This is achieved by measuring the post-excitation population in the $|5s, F=2, m_F=-2 \rangle$ initial state as a function of the excitation time, to which we fit an exponential to extract the pumping rate. We obtain the resonant rate $R_{0}$ due to the Rydberg $s$ state by subtracting the off-resonant $5s-5p$ optical pumping rate. The measurements presented in the main text are done with two different fractional densities: $f=25\%$ and $f=100\%$.

\bibliography{CumulantSPRefs}

\end{document}